\documentclass[aps,11pt,preprintnumbers,axodraw,nofootinbib]{revtex4}
\pdfoutput=1
\usepackage{epsfig}
\usepackage{amsmath}
\usepackage{bm}
\usepackage{times}
\usepackage{graphicx}
\usepackage{color}
\usepackage{slashed}

\def\nn{\nonumber}
\def\bea{\begin{eqnarray}}
\def\eea{\end{eqnarray}}
\def\ba{\begin{eqnarray}}
\def\ea{\end{eqnarray}}
\def\be{\begin{equation}}
\def\ee{\end{equation}}

\begin{document}
\preprint{Published in JHEP}

\title{Prospects and Constraints for Vector-like MFV Matter at LHC}

\author{Jonathan M. Arnold$^{1}$, Bartosz Fornal$^{1}$ and
Michael Trott$^{2}$ \\
${}^1$ California Institute of Technology, Pasadena, CA 91125, USA\\
${}^2$ Perimeter Institute for Theoretical Physics, Waterloo ON,
N2L 2Y5, Canada.\\
}
\date{\today}

\begin{abstract}
We examine the prospects for LHC discovery of $\rm SU(2)_L$ singlet vector-like quarks that obey Minimal Flavour Violation (MFV)
and are consistent with lower energy phenomenology.
We study models where the vector-like quarks have the same quantum numbers as $u_R$ or $d_R$,
allowing mixing, which generally leads to significant low energy constraints.
We find that
at leading order in the MFV expansion
there are two naturally phenomenologically viable MFV models of this type
when the Weyl spinor components of the vector-like quarks are flavour triplets.
We examine direct production bounds, flavour and electroweak precision data constraints for these models and
determine the cross section for allowed values of the model parameters at LHC. For the models we identify as
naturally phenomenologically viable, large amounts of parameter space afford a significant early discovery reach at LHC
while being consistent with lower energy phenomenology.

\end{abstract}
\maketitle

\bigskip

\section{Introduction}

In addition to the known standard model (SM) quarks, models beyond the SM generally
include new matter content. Such extended matter content is strongly constrained by
the requirement of gauge anomaly cancellation \cite{Georgi:1972bb}.
In recent years, vector-like quarks (for reviews see \cite{Frampton:1999xi,AguilarSaavedra:2002kr})
have been studied since the possibility of vector-like $\rm SU(2)_L$ singlet quarks
is a particularly simple and minimal anomaly free extension of the SM.
From a top-down model building perspective, vector-like matter of this form is also interesting.
Extensions of the SM involving new vector-like matter have been explored in the context of
SUSY to address the little hierarchy problem \cite{Barger:2006fm,Babu:2008ge,
Liu:2009cc,Graham:2009gy}, and triplets of vector-like quarks can also appear in grand unified models
and string compactifications that embed the SM into an $\rm E_6$ group \cite{Rosner:1985hx}.

The current phenomenological constraints on vector-like matter come
from direct production bounds at the Tevatron and flavour physics.
Tevatron studies  exclude a heavy $t'$ with SM like couplings at $\rm 95 \%$ CL
up to $335 \, {\rm GeV}$ \cite{CDFtquark} and a heavy $b'$ with SM like couplings at $\rm 95 \%$ CL
up to $338 \, {\rm GeV}$ \cite{Aaltonen:2007je}.
Flavour constraints are also significant  for vector-like quarks. The mixing of $n$ vector-like
quarks with the SM quarks causes the $3 \times 3$ SM CKM matrix to become non-unitary as it is
extended to a $(3+n) \times (3+n)$ unitary matrix. Non-unitarity of this form is tightly constrained.  By mixing with the vector-like quarks, effective tree level flavour changing
neutral currents (FCNC) could also be induced for the SM quarks (see \cite{AguilarSaavedra:2002kr} for recent studies).
Although flavour constraints are stringent due to the lack of $\it any$ clearly statistically significant evidence of non-SM FCNC's (see, e.g., \cite{Isidori:2010kg} for a survey), these constraints do not provide a direct mass bound, as the
mixing factors are generally unknown and can be chosen to be small.

Thus, it is interesting to examine vector-like quark models within the framework
of Minimal Flavour Violation (MFV) \cite{Chivukula:1987py,D'Ambrosio:2002ex,Feldmann:2008ja,Kagan:2009bn,Feldmann:2009dc}
where new physics effects are
flavour conserving apart from the effects of flavour breaking by insertions of Yukawa matrices, as in the SM.
MFV is also predictive in terms of the allowed representations and couplings of new physics.
For example, it has been shown that only a single flavour singlet scalar
representation is allowed by MFV besides the Higgs \cite{Manohar} (see \cite{Kim:2008bx,Idilbi:2009cc,Burgess:2009wm,Fornal:2010ac} for recent studies).
When the scalar is allowed to transform under the flavour symmetry
more representations are allowed \cite{Arnold:2009ay}, but MFV remains a predictive formalism that constrains the allowed mass spectra
and couplings. Similarly, for the vector quark models we study, the strength of the vector-quark quark mixings are no longer uncorrelated free parameters
in the theory due to MFV.

In this paper we examine
vector-like quark models whose Weyl spinors transform as flavour triplets when MFV is an approximate global symmetry.
We will show that
at leading order in the MFV expansion
only two models exist that are naturally
phenomenologically viable; one with a triplet of charge $Q = -1/3$ vector-like quarks and one with a triplet of charge $Q = 2/3$ vector-like quarks.  By naturally phenomenologically viable we mean that
the new flavour changing effects in the model are suppressed, in agreement with experiment,
not simply through the choice of parameters but in a way that follows from the group structure and representation content of the theory.
This flavour naturalness criteria is a generalization of the
Glashow-Weinberg criteria for natural FCNC suppression \cite{Glashow:1976nt} and is
not guaranteed by MFV alone.
The continuing success of the SM in the flavour physics program, with ever more flavour changing decays and flavour oscillation measurements probing the weak scale and consistent with the SM,
strongly implies that models which satisfy this naturalness criteria are promising models to search for in the LHC era.

The outline of this paper is as follows. In Section II we present our procedure for
solving the mass spectrum of the models and determine the two
models that are naturally viable. We then determine the allowed parameter space
considering relevant collider,  flavour and electroweak precision data (EWPD) constraints in Section III.
We then turn to the expected cross sections at LHC for the allowed remaining parameter space of these models in Section IV.
Finally, we comment on similar flavour triplet MFV models of this form when $m_V  \gtrsim \rm TeV$.

\section{MFV vector-like quark models}

The SM has a global $\rm U(3)^5$ flavour symmetry that is only broken by the SM Yukawa interactions.
Decomposing the abelian quark subgroup of this symmetry as
\ba
G_{F} = \rm SU(3)_u \otimes SU(3)_d \otimes SU(3)_Q\ ,
\ea
one can formally restore this subgroup of the flavour symmetry by treating the Yukawa matrices as
spurions that transform as
\ba
g_d \sim (1,\bar{3},3)\ , \quad \, \quad g_u \sim (\bar{3},1,3)\ ,
\ea
such that MFV forbids FCNC's at tree level.
The flavour symmetry is broken by insertions of the
spurions $g_u \, g_u^\dagger$ or $g_u^\dagger \, g_u$ when $\rm SU(3)_{U_R}$ or  $\rm SU(3)_{Q_L}$ indices are contracted.
We use notation where each insertion is
accompanied by (the unknown constant) $\eta_i$ and choose $\eta_i \sim \eta \ll 1$.
Except for the top, all Yukawas are already small so the symmetry breaking effects are highly suppressed.
For the top, our perturbative expansion in $\eta$ is appropriate under this assumption.
The naturalness of this assumption depends on the UV completion of the models under study
and specifying a UV completion is beyond the scope of this work.
For approaches to MFV without this assumption see \cite{Feldmann:2008ja,Kagan:2009bn,Feldmann:2009dc}.
In our analysis, due to this assumption, the effects of mixing are
the dominant constraint on the mass scale of the model.

In this paper, we are considering models whose left- and right-handed
(Weyl spinor) components transform as flavour triplets.  Similar MFV models have been examined before in both the quark and lepton
sector \cite{Grossman:2007bd,Gross:2010ce}. Both of these works assume that the
bare mass parameters are far larger than the electroweak scale $v$, which can be appropriate for $\geq$ TeV scale vector-like quark masses.
In this paper, we directly solve the models
at leading order in the MFV expansion
without this assumption and study the constraints and prospects for these models in much greater detail.
This turns out to be instructive as we argue it singles out two models that are naturally phenomenologically viable.
We will discuss our approach in some detail for an extension of the SM with vector-like down quarks.
The discussion for vector-like up quarks is similar.

\subsection{Vector-like down quark models}

The Lagrangian including the right- and left-handed chiral fields of the vector-like quarks is the following
\begin{eqnarray}\label{Lag}
{\cal L}^d&=& \bar{Q}_Li\slashed{D}Q_L+ \bar{d}_Ri\slashed{D}d_R+ \bar{V}_L^d i\slashed{D}V_L^d+ \bar{V}_R^d i\slashed{D}V_R^d\nn\\
 &&+\left[\kappa_1^d m_1^d \, \bar{V}_L^d \, d_R + \kappa_2^d m_2^d\bar{V}_L^d V_R^d+ \kappa_3^d  \, \left(\tfrac{\sqrt{2} \, m_3^d}{v} \right) \, \bar{Q}_L H V_R^d + g_d \bar{Q}_L H d_R +\rm h.c.\right]\ .
\end{eqnarray}
The transformations allowed under $G_F$ are listed in Table I for the new couplings and fields. Note that the couplings $\kappa_i^d$ are matrices whose transformation law depends on the model and is such that the Lagrangian is invariant under $G_F$. MFV requires that these matrices be
Yukawas or identity matrices.
The factors $m_i^d$ set the scale of the couplings.
The fields $V_R^d$, $V_L^d$ are triplets under $\rm SU(3)_c$, singlets under $\rm SU(2)_L$ and have hypercharge $-1/3$.
As in the SM, when MFV is imposed, $Q_L$ transforms as $(1,1,3)$ while $d_R$ transforms as $(1,3,1)$ under $G_{F}$.

\vspace{0.8cm}
\begin{table}[ht]
\begin{center}
\begin{tabular}[t]{|c|c|c|c|c|c|c|}
  \hline
  \hline
    & \multicolumn{5}{|c|}{$SU(3)_{U_R} \times SU(3)_{D_R} \times SU(3)_{Q_L}$} \\ \hline
    model & $\kappa_1^d$  & $\kappa_2^d$  & $\kappa_3^d$ & $V_{L}^d$ & $V_{R}^d$  \\  \hline
I &  \,  0 \, & \, (1,1,1)\,   & \, (1,$\bar{3}$,3)\,   & \,(1,3,1)\, & \,(1,3,1)\, \\
II &   (1,$\bar{3}$,3) & \, (1,1,1)\,  & (1,1,1)  & (1,1,3) & (1,1,3) \\
III &   0 & (1,$\bar{3}$,3) & \, (1,$\bar{3}$,3)\,   & (1,1,3) & (1,3,1) \\
IV &   (1,1,1)* & (1,3,$\bar{3}$) & \, (1,1,1)\, & (1,3,1) & (1,1,3) \\
V &   (1,$\bar{3}$,3) & ($\bar{3}$,1,3) & \, ($\bar{3}$,1,3)\,  & (1,1,3) & (3,1,1) \\ \hline
\hline
\end{tabular}
\end{center}
\caption{Representations of $V_L^d, V_R^d$ allowed by MFV. We have used the fact that in models $\rm I,III $
the $ V_R^d$ has the same transformation as $d_R$ under the
flavour and gauge groups to rotate to a basis in
$V_R^D - D_R$ space where the mixing term vanishes \cite{Grossman:2007bd}.
We also note that model IV predicts non-hierarchical corrections to SM masses due to mixing. These corrections require extra tuning to cancel against the SM masses when $m_1^d, m_2^d \gg v$.
This was indicated with a star appended to the relevant coupling.}
\label{int}
\end{table}

\subsection{Solving vector-like down quark models}
We now diagonalize the Lagrangians corresponding to the models listed in Table I.
There are two forms of mixing now to contend with when vector-like quarks are added to the
SM;  mixing that in the SM corresponds to the transformation between the
weak eigenstates and the mass eigenstates, and mixing between the vector-like quarks and
the SM quarks. We diagonalize in a two-stage procedure.
The fields that correspond to the initial Lagrangian terms in Eq. (\ref{Lag}) are designated as the
unprimed fields.
We introduce unitary transformations to primed fields (which would be mass eigenstates in the SM if no vector-like quarks were present) and we
introduce analogous unitary transformations for the vector-like quarks:
\bea
d_R &=&  \mathcal{U}(d,R) \, d_R'\ ,  \quad \quad d_L =  \mathcal{U}(d,L) \, d_L'\ ,\nn \\
u_R &=&  \mathcal{U}(u,R) \, u_R'\ ,  \quad \quad u_L =  \mathcal{U}(u,L) \, d_L'\ ,\nn \\
V^d_R &=&  \mathcal{U}(V,R) \, V^{d\,\prime}_R\ , \quad \ \  V^d_L =  \mathcal{U}(V,L) \, V^{d\,\prime}_L\ .
\eea
We choose them in such a way that the initial Yukawa matrices
are diagonalized by taking\footnote{Although $\mathcal{M}^0_i$ is a real diagonal matrix and $v$ is the vacuum expectation value of the real component of the Higgs doublet,
until the further mixing effects between the vector-like quarks and the SM quarks are taken into account
one cannot identify the elements in this matrix in terms of the physical quark masses as yet.}
\bea
 \mathcal{U}^\dagger(d,L)  \, g_d \,  \mathcal{U}(d,R) = \frac{\sqrt{2} \mathcal{M}^0_d}{v}\ , \quad  \mathcal{U}^\dagger(u,L)  \, g_u \,  \mathcal{U}(u,R) = \frac{\sqrt{2} \mathcal{M}^0_u}{v}\ .
\eea
We write the Lagrangian in terms of the primed fields and the mass matrix $M_d$
\bea
\mathcal{L}_{m} = \left(
\begin{array}{c}
\bar{d'}_L \\
\bar{V}^{d\,\prime}_L
\end{array}
\right)^T
M_d
\left(
\begin{array}{c}
d'_R \\
V^{d\,\prime}_R
\end{array}
\right)\ ,
\eea
where
\vspace{-7pt}
\bea
\label{diag}
M_d=\left(
\begin{array}{cc}
\mathcal{M}^0_d\  &\  m_3^d \,  \mathcal{U}^\dagger(d,L) \,  \kappa_3^d  \,  \mathcal{U}(V,R) \\
m_1^d \,  \mathcal{U}^\dagger(V,L) \, \kappa_1^d \,  \mathcal{U}(d,R) \ & \ m_2^d  \,  \mathcal{U}^\dagger(V,L) \, \kappa_2^d   \,  \mathcal{U}(V,R) \\
\end{array}
\right)\ .
\eea

We designate the true mass eigenstates of the SM quarks and the vector-like quarks as double-primed fields.
The primed basis is now purely unphysical for the vector-like quarks;  it does
not correspond to the eigenstates of any interaction. Thus we are free to choose $\, \mathcal{U}(V,R),  \,\mathcal{U}(V,L) $ in order to diagonalize each
 $3 \times 3$ block of the Hermitian matrices in $M_d$ as
any rotation of the vector-like quark field basis is allowed.  This can be accomplished (up to insertions of $V_{CKM}$) by choosing the $\mathcal{U}(V,R/L)$ to be the appropriate
$\mathcal{U}(u/d,R/L)$ or $\bf 1$ as MFV forces the $\kappa_i^d$ to be (undiagonalized) Yukawa matrices
or the identity matrix.

We determine the $M_d  M_d^\dagger$ and $M_d^\dagger  M_d$ matrices in terms of  the SM parameters in $\mathcal{M}^0_d, \mathcal{M}^0_u$, and $V_{CKM}$.
The  $6 \times 6$ matrix $M_d$ can always be diagonalized by
introducing appropriate rotations of the primed fields to the mass eigenstate basis:
\begin{align}
\left(
\begin{array}{c}
d''_L \\
V^{d\,\prime\prime}_L
\end{array}
\right)
&=U
\left(
\begin{array}{c}
d'_L \\
V^{d\,\prime}_L
\end{array}
\right)\ ,\ \ \
\left(
\begin{array}{c}
d''_R \\
V^{d\,\prime\prime}_R
\end{array}
\right)
=W
\left(
\begin{array}{c}
d'_R \\
V^{d\,\prime}_R
\end{array}
\right)\ ,
\end{align}
where $U, W$ are the unitary transformations diagonalizing $M_d M_d^\dagger$, $M_d^\dagger M_d$, respectively,
and $D=U M_d W^\dagger$ is diagonal. Writing these transformations as
 \begin{align}
U=\left(
    \begin{array}{cc}
      u_{11} & u_{12} \\
      u_{21} & u_{22} \\
    \end{array}
  \right)\ , \ \ \ W=\left(
    \begin{array}{cc}
      w_{11} & w_{12} \\
      w_{21} & w_{22} \\
    \end{array}
  \right)\ ,
\end{align}
where $u_{ij}, w_{ij}$ are $3\times3$ diagonal matrices, the general form for the kinetic terms is
\bea
\mathcal{L}_{\rm kin}= \mathcal{L}^{SM}_{\rm kin} &-& \left(u_{11}^2-1\right)\tfrac{\sqrt{g_1^2+g_2^2}}{2}\, \bar{d}_L''\slashed{Z}d_L''
+\bar{V}_R''\left[i\slashed{\partial}+\tfrac{g_1 \sin{\theta_W}}{3}\slashed{Z}-\tfrac{g_2 \sin{\theta_W}}{3}\slashed{A}\right]V_R''\nn\\
&+& \bar{V}_L''\left[i\slashed{\partial}+\left(-u_{21}^2\tfrac{\sqrt{g_1^2+g_2^2}}{2}+\tfrac{g_1 \sin{\theta_W}}{3}\right)\slashed{Z}-\tfrac{g_2 \sin{\theta_W}}{3}\slashed{A}\right]V_L''\nn\\
&+&\tfrac{g_2}{\sqrt{2}}\bar{u}_L''V_{CKM}\slashed{W}^+\left[(u_{11}-1)d_L''+u_{21}V_L''\right] +
\tfrac{g_2}{\sqrt{2}}\left[(u_{11}-1)\bar{d}_L''+u_{21}\bar{V}_L''\right]V^\dagger_{CKM}\slashed{W}^-u_L''\,\nn \\
&-&u_{11}u_{21}\tfrac{\sqrt{g_1^2+g_2^2}}{2}\left(\bar{d}_L''\slashed{Z}V_L''+\bar{V}_L''\slashed{Z}d_L''\right)\ ,
\eea
which will be used to derive the constraints for the vector-like quark models.
The up quarks do not require any further rotation so that $u'_{L/R} =u''_{L/R} $.

Note that the SM CKM matrix is defined as $V_{CKM}=  \mathcal{U}^\dagger(u,L) \, \mathcal{U}(d,L)$
and
\bea
\left(
\begin{array}{c}
u_L \\
d_L
\end{array}
\right) &=&  \mathcal{U}(u,L)
\left(
\begin{array}{c}
u_L' \\
V_{CKM} d_L'
\end{array}
\right).
\eea
The CKM matrix in these models corresponds to the difference in rotating the up-type quarks in the $\rm SU(2)_L$ doublet compared to the down-type quarks (which are an admixture of $d_L''$ and $V_L''$)
from the unprimed weak basis to the double-primed mass basis. The resulting $3 \times 3$ SM CKM matrix is now non-unitary due to the correction of the $\bar{u}_L \, W^+ \, d_L + h.c.$ coupling that depends on $(u_{11} -1)$.
An advantage of our approach to diagonalization is that the resulting non-unitary $3 \times 3$ SM CKM matrix has a clear decoupling limit.

Using this procedure to solve the Table I models (and a similar procedure for the vector-like up quark models) singles out one model in each case that is more naturally phenomenologically viable (at leading order in the MFV expansion).
We now discuss in some detail this vector-like down quark model before turning to the effects of flavour breaking.
\vspace{-0.4cm}
\subsubsection{A viable vector-like down quark model}
\vspace{-0.2cm}
In model $\rm I$ we have  $\kappa_1^d = 0,\kappa_2^d ={\bf 1},\kappa_3^d  = g_d$. Without loss of generality we have chosen $\mathcal{U}(V,L) = \mathcal{U}(V,R) = \mathcal{U}(d,R)$ to work in a basis where the $3 \times 3$ subblocks of
Eq. (\ref{diag}) are diagonal.
We find the diagonal form of the squared mass matrix
\vspace{-0.2cm}
\bea
D^2  =  \left(
\begin{array}{cc}
\left(\mathcal{M}^0_d\right)^2 & 0 \\
0 & (m^d_2)^2\left( 1 + \epsilon_i^2\right) \\
\end{array}
\right),
\eea
\vspace{-0.2cm}
where $\epsilon_i  = \sqrt{2} \, \frac{m^d_3}{m^d_2} \frac{(\mathcal{M}^0_d)_i}{v}$ and
we neglected terms  $\mathcal{O}(\epsilon_i^4)$.
From the normalized eigenvector matrices one identifies the $U$ and $W$ transformations as
\bea
U = \left(\begin{array}{cc}
{\bf 1} - \tfrac{\epsilon_i^2}{2}   & - \epsilon_i \\
\epsilon_i & {\bf 1}  - \tfrac{\epsilon_i^2}{2} \\
\end{array}
\right)\ ,
\quad
W = \left(\begin{array}{cc}
\ {\bf 1}\  &  \ - \tfrac{\epsilon_i^2}{\sqrt{2} \, \xi_3}  \  \\
\  \tfrac{\epsilon_i^2}{\sqrt{2} \, \xi_3} \  & \ {\bf 1}\   \\
\end{array}
\right)\ ,
\eea
where we neglected terms  $\mathcal{O}(\epsilon_i^4)$. Note that $\xi_i = m^d_i/v$.
We do not assume that $v \ll m^d_{1,2,3}$ or a large hierarchy of masses $m^d_3 \gg m^d_2$. We are interested in solutions where the elements of the diagonalized $\mathcal{M}^0_d$ (given by $m^0_i$) are such that $m^0_i\ll v$ thus $\epsilon_i \ll 1$.
We note that no further source of CP violation is present
due to the final rotation that diagonalizes the masses, and the SM masses are identified identically with $m_i^0$ in this model.
We also note that rotations of $V^d_{R,L}$ can eliminate explicit CP violating phases from the new Lagrangian terms in Eq. (\ref{Lag}) in model I, and that
here the indices on the $\epsilon_i$ are treated as labels coincident with the flavour index, not true flavour indices, i.e. they are not contracted.

This model has a number of features that distinguish it from the remaining models of Table I
and make it naturally phenomenologically viable:
\begin{itemize}
\item [(i)] The non-unitarity of the CKM matrix is due to a correction of the form
\bea
V_{CKM} = V_{CKM}^{SM}\left[{\bf 1} + \left(\tfrac{m^d_3}{m^d_2}\right)^2 \tfrac{(\mathcal{M}^0_d)^2_i}{v^2} \right]\ .
\eea
This leads to small deviations from the SM  CKM;  the largest deviations are in
$V_{tb}, V_{ub}, V_{cb}$, which are less precisely measured, and are proportional to $m_b^2/v^2$. As we discuss in Section \ref{flavour}, the deviation
from unitarity, being contributed to only the diagonal terms of the square of the CKM matrix, removes contributions to
meson mixing observables.

\item [(ii)] The modification of the SM couplings of the down quarks due to mixing with the vector-like quarks
is proportional to
\vspace{-0.5cm}
\bea
\delta (\bar{d}_L''\slashed{Z}d_L'') \propto \tfrac{\sqrt{g_1^2+g_2^2}}{2} \,  \left(\tfrac{m^d_3}{m^d_2}\right)^2  \frac{(\mathcal{M}^0_d)^2_i}{v^2}\ .
\eea
MFV does not guarantee small mixing effects on the SM quark couplings to the $W^\pm,Z$.
This model, and a similar vector-like up quark model, do have small mixing effects without the mass scale of the
vector-like quarks being $m_V \gg v$.
\end{itemize}

The final phenomenologically relevant couplings are those involving the Higgs
that are generated from terms initially involving vector-like quarks. Up to order $\mathcal{O}(\epsilon_i)$,
\bea
\mathcal{L}_{Vh} &=& \epsilon_i  \, \xi_2 \, (\bar{d}''_L)_i \, h \, (V^{d\,\prime\prime}_R)^i  +  \epsilon_i  \, \xi_2 \,  (\bar{V}^{d\,\prime\prime}_R)_i \, h \, (d''_L)^i\ ,
\eea
where $\,\xi_{2}=\tfrac{m^d_{2}}{v}$. The SM Yukawa coupling receives a small correction $\mathcal{O}(\epsilon_i^2)$ which we have neglected.
We have also neglected the effects of spurion breaking of the $ {\rm U}(3)^5$ flavour symmetry but have explicitly included the
flavour indices. We now determine the flavour breaking effects up to leading order in $\eta$ and the dominant decay widths of this model.
\vspace{-0.2cm}
\subsubsection{Flavour breaking and decays in model I}
\vspace{-0.1cm}
The dominant flavour breaking effects come from insertions of the $g_u^\dagger \, g_u$
or $g_u \, g_u^\dagger$ spurions. The kinetic terms have suppressed flavour indices that one can contract with spurion insertions.
Canonically normalizing the fields after such insertions ensures that only interaction terms in the Lagrangian that are not bilinear in $Q_L, u_R$ or $d_R$ will
receive corrections of this form. Thus, the only terms that receive flavour breaking corrections are
\vspace{-5mm}
\bea
\delta \mathcal{L} = \epsilon_i  \, f_1(\eta) \, \xi_2 \, (\bar{d}''_L)_i \, h \, (V^{d\,\prime\prime}_R)^i  +  \epsilon_i  \, f_1(\eta) \, \xi_2 \,  (\bar{V}^{d\,\prime\prime}_R)_i \, h \, (d''_L)^i\ .
\eea
Taking into account the canonical rescaling of the $Q_L$ field one finds $ f_1(\eta) = 1 + \eta  \, \tfrac{m_t^2}{v^2} \, \delta_{i,3}$.
These flavour symmetry breaking terms will not be the focus of our analysis since they do not provide clearly dominant
constraints on the mass scale of the theory (they depend on the unknown symmetry breaking parameter $\eta$ and we assumed $\eta \ll 1$).   Our focus will be on the larger flavour diagonal
mixing effects with the SM quarks that lead to significant mass constraints independent of $\eta$.

The $V_R^d$ dominantly decay through Higgs interactions while
the $V^d_L$ dominantly decay through charged current interactions to SM up quarks which can afford correlated collider signatures. The largest decay widths are given by
\bea
\label{gam}
\Gamma((V^d_L)^j \rightarrow Z\, (d_L)^j) &=& \frac{(g_1^2 + g_2^2) \, \epsilon_j^2}{128 \, \pi} \frac{(m^d_j)^3}{m_Z^2} \left(1 - \frac{m_Z^2}{(m^d_j)^2} \right)^2  \left(1 + 2 \, \frac{m_Z^2}{(m^d_j)^2} \right),  \\
\Gamma((V^d_L)^j \rightarrow W^- \, (u_L)_i) &=& \frac{g_2^2 \,  \epsilon_j^2 |(V_{CKM})^j_i|^2}{64 \, \pi} \, \frac{(m^d_j)^3}{m_W^2} \, \left(1 - \frac{m_W^2}{(m^d_j)^2} \right)^2 \left(1 + 2 \, \frac{m_W^2}{(m^d_j)^2} \right), \\
\Gamma((V^d_R)^j \rightarrow h \, (d_R)^j) &=& \frac{\epsilon_j^2 \, \xi_2^2 \, f_1(\eta)^2}{32 \, \pi} \, m^d_j \, \left(1 - \frac{m_h^2}{(m^d_j)^2} \right)^2\ .
\eea


\subsection{Vector-like up quarks}

The Lagrangian including the fields $V_R^u$, $V_L^u$ is
\bea\label{lag2}
{\cal L}^u&=& \bar{Q}_Li\slashed{D}Q_L+ \bar{u}_Ri\slashed{D}u_R+ \bar{V}_L^u i\slashed{D}V_L^u+ \bar{V}_R^u i\slashed{D}V_R^u\nn\\
&+&\left[\kappa_1^u m_1^u\bar{V}_L^u u_R+\kappa_2^u m_2^u\bar{V}_L^u V_R^u+ \kappa_3^u \, \left(\tfrac{\sqrt{2} \, m_3^u}{v} \right)\,\bar{Q}_L H^{\dagger} V_R^u + g_u \bar{Q}_L H^{\dagger} u_R +\rm h.c.\right] \ .
\eea
The fields $V_R^u$, $V_L^u$ are triplets under $\rm SU(3)_c$, singlets under $\rm SU(2)_{\rm L}$ and have hypercharge $+2/3$. The allowed representations are listed in Table II.
\vspace{2mm}
\begin{table}[h]
\begin{center}
\begin{tabular}[t]{|c|c|c|c|c|c|}
  \hline
  \hline
    & \multicolumn{5}{|c|}{$SU(3)_{U_R} \times SU(3)_{D_R} \times SU(3)_{Q_L}$} \\ \hline
    model & $\kappa_1^u$  & $\kappa_2^u$ & $\kappa_3^u$& $V_{L}^u$ & $V_{R}^u$ \\  \hline
VI &  \, 0   \, & \, (1,1,1)\,   & \,($\bar{3}$,1,3) \,& \,(3,1,1)\, & \,(3,1,1)\, \\
VII &   ($\bar{3}$,1,3) & (1,1,1) & (1,1,1) & (1,1,3) & (1,1,3) \\
VIII &  0  & ($\bar{3}$,1,3) &  ($\bar{3}$,1,3) & (1,1,3) & (3,1,1) \\
IX &   (1,1,1)* & (3,1,$\bar{3}$) & (1,1,1) & (3,1,1) & (1,1,3) \\
X &  ($\bar{3}$,1,3) & (1,$\bar{3}$,3) & (1,$\bar{3}$,3) & (1,1,3) & (1,3,1) \\ \hline
\hline
\end{tabular}
\end{center}
\caption{Representations of $V_L^u, V_R^u$. Model IX  predicts non-hierarchical corrections to SM quark masses, this was again indicated with a star.}
\end{table}

\subsection{Vector-like up quark Lagrangian construction}\label{vups}

If we assume a mild hierarchy $m_2 > m_3$ such that $\epsilon_i$ is still a good expansion parameter for the top quark, the vector-like up quark models are nearly identical to the corresponding down-type models except for modifications\footnote{One replaces  e.g., $d_R \rightarrow u_R,~~ \mathcal{U}(d,L) \rightarrow \mathcal{U}(u,L), ~V_{CKM}\rightarrow V_{CKM}^\dagger$, and $\mathcal{M}_d^0 \rightarrow \mathcal{M}_u^0\ .$~} in the kinetic Lagrangian corresponding to having $Q=+\frac{2}{3}$ and $T^3_{u_L}=+\frac{1}{2}$. The kinetic part of the Lagrangian is
\bea
\mathcal{L}_{\rm kin}= \mathcal{L}^{SM}_{\rm kin} &+& \left(u_{11}^2-1\right)\tfrac{\sqrt{g_1^2+g_2^2}}{2}\, \bar{u}_L''\slashed{Z} \, u_L''
+\bar{V}_R''\left[i\slashed{\partial}-\tfrac{2 \, g_1 \sin{\theta_W}}{3}\slashed{Z}+\tfrac{2 \, g_2 \sin{\theta_W}}{3}\slashed{A}\right]V_R''\nn\\
&+& \bar{V}_L''\left[i\slashed{\partial}+\left(u_{21}^2\tfrac{\sqrt{g_1^2+g_2^2}}{2}-\tfrac{2 \, g_1 \sin{\theta_W}}{3}\right)\slashed{Z}+\tfrac{2 \, g_2 \sin{\theta_W}}{3}\slashed{A}\right]V_L''\nn\\
&+&\tfrac{g_2}{\sqrt{2}}\left[(u_{11}-1)\bar{u}_L''+u_{21}\bar{V}_L''\right] \, V_{CKM}\slashed{W}^+ d_L'' +
\tfrac{g_2}{\sqrt{2}} \, \bar{d}_L'' \, V^\dagger_{CKM}\slashed{W}^-\left[(u_{11}-1){u}_L''+u_{21}{V}_L''\right]\,\nn \\
&+&u_{11}u_{21}\tfrac{\sqrt{g_1^2+g_2^2}}{2}\left(\bar{u}_L''\slashed{Z}V_L''+\bar{V}_L''\slashed{Z}u_L''\right) \ .
\eea
As before, we must allow for insertions of both $g_u \,g_u^\dagger$ and $g_u^\dagger \, g_u$ anywhere we have contractions of $\rm SU(3)_{{\rm u}_R}$ and $\rm SU(3)_{Q_L}$ indices, respectively.
Thus, insertions of $f(\eta)$ occur more frequently in the up-type models and provide non-trivial differences between the phenomenology derived in the down-type models.

\subsubsection{A viable MFV vector-like up quark model}

The viable MFV up quark model is model $\rm VI$. The analysis of this model
proceeds as in Section II B 1 with the appropriate replacements.
The only difference in the analysis comes from the flavour breaking effects.
Similarly to model I, we have the following Higgs interaction terms
\bea
\mathcal{L}_{Vh} &=& \epsilon_i  \, \xi_2 \,  f_2(\eta) \, (\bar{u}''_L)_i \, h \, (V^{d\,\prime\prime}_R)^i  +  \epsilon_i  \, \xi_2 \,   f_2(\eta) \, (\bar{V}^{d\,\prime\prime}_R)_i \, h \, (u''_L)^i\ .
\eea
The flavour breaking corrections are given by $f_2(\eta) = 1 + 3 \, \eta  \, \tfrac{m_t^2}{v^2} \, \delta_{i,3}$
and the partial widths are obtained directly from Eqs. (18)-(20) with the appropriate replacements.

\subsection{Remaining vector-like flavour triplet models}

We have emphasized that the models which we study in detail are naturally phenomenologically viable under the assumption $\eta \ll 1$.
We have also solved the remaining models in Tables I, II  and
summarize their mass spectrum and mixing angles in the Appendix. These other models become more viable
as the mass scale $m_V$ increases and mixing effects decrease in the decoupling limit, along with LHC discovery potential.
One can always choose the parameters in  Eqs. (\ref{Lag}), (\ref{lag2}) to reduce mixing effects for the models in the Appendix, violating our naturalness
criteria.  Such a tuning is also not protected by any symmetry against radiative corrections.
In the remainder of the paper, we will focus our attention on the two naturally phenomenologically viable models.

\section{Constraints on vector-like quark models}

We will study EWPD constraints, flavour physics constraints
and current collider constraints on models I, VI. We begin with
the current collider constraints. In each case we analyze the relevant constraint
assuming that the SM is extended by model I $\it or$ model VI.

\subsection{Collider constraints}\label{collider}

\subsubsection{Vector-like down quarks}
For model I we use the model-independent results from \cite{Aaltonen:2007je} constraining deviations from SM predictions of $\ p \, \bar{p} \rightarrow Z \, + {(\geq 3 j)}$ at the Tevatron.
The addition of singlet vector-like quarks does not supply a positive contribution to the $\rm T$ EWPD
parameter that can raise the fitted Higgs mass value. As the decays $h \rightarrow ZZ$ are highly suppressed for the best fit Higgs mass values (recent studies find $ M_H = 96^{+29}_{-24} {\,\rm GeV} $ \cite{Erler:2009jh}) only
the pair production of $V_L$  will yield significant branchings to $Z \, + (\geq 3 j)$. Due to this we rescale the simulated production cross section
(which has appropriate acceptance and phase space cuts for the analysis \cite{Aaltonen:2007je}) shown in Fig. 38 of this study by $\frac{1}{2}$.

We also rescale to correct for the branching ratio to the constrained final states. The experimental study assumed  ${\rm BR}(b' \rightarrow  Z \, + (\geq 3 j)) = 1$.
The effective branching ratio for the produced $V^d_L$ to produce the signal ($\beta_{\rm eff}$) is
\bea
\beta_{\rm eff} &\simeq& \, {\rm BR}(V_L^j \rightarrow Z  \, d_L^j) \times  {\rm BR}(\bar{V}_L^j \rightarrow W^+  \, \bar{u}_L^i) \times {\rm BR}( W^+  \rightarrow {\rm hadrons}) \nn \\
&\, &+\ \,{\rm BR}(\bar{V}_L^j \rightarrow Z  \, \bar{d}_L^j) \times  {\rm BR}(V_L^j \rightarrow W^-  \, u_L^i) \times {\rm BR}( W^-  \rightarrow {\rm hadrons}) \nn \\
& \,& + \   \, {\rm BR}(V_L^j \rightarrow Z  \, d_L^j) \times{\rm BR}(\bar{V}_L^j \rightarrow Z \, \bar{d}_L^j) \times {\rm BR}( Z  \rightarrow {\rm hadrons})\ .
\eea
From the PDG we determine ${\rm BR}( Z / W^\pm  \rightarrow {\rm hadrons})$ and we assume the produced hadrons initiate jets that will pass the triggers of
 \cite{Aaltonen:2007je}. From Eqs. (18-20) we also determine
\bea
 {\rm BR}(V_L^j \rightarrow Z  \, d_L^j)  &\simeq& \frac{(g_1^2+g_2^2) \, m_W^2}{(g_1^2+g_2^2)m_W^2+2g_2^2 m_Z^2}\ , \\
 {\rm BR}(V_L^j \rightarrow W^-  \, u_L^i)  &\simeq& \frac{2  g_2^2  m_Z^2}{(g_1^2+g_2^2)m_W^2+2g_2^2 m_Z^2}\ ,
 \eea
where we have neglected corrections of  $\mathcal{O}(m_{W}^2/m_V^2)$, $\mathcal{O}(m_{Z}^2/m_V^2)$.
Only retaining contributions that are not CKM or light quark Yukawa suppressed we find $\beta_{\rm eff} \simeq 0.38$
using couplings and masses defined at $\mu = m_Z$. Rescaling the LO $\sigma(p \, \bar{p} \rightarrow b' \, \bar{b}')$
curve of \cite{Aaltonen:2007je} in Fig. 38 by $\tfrac{N_f}{2}\beta_{\rm eff}$, we obtain a 95\% CL lower bound for the vector-like down mass of 200 GeV for model I.

Recent studies \cite{CDFbquark} searching for $b'  \, \bar{b'}  \rightarrow (t W^{\mp}) \, (\bar{t} W^{\pm})$ improve this bound, assuming the decay products pass the corresponding triggers.
To use this study the effective branching ratio is given by
\bea
\beta_{\rm eff} &\simeq& {\rm BR}(V_L^3 \rightarrow W^-  \, t_L) \times {\rm BR}(\bar{V}_L^3 \rightarrow W^+  \, \bar{t}_L) \simeq 0.44\ .
\eea
Rescaling the results in Table II of  \cite{CDFbquark} by $\tfrac{1}{2}\beta_{\rm eff}$ we obtain a 95\% CL lower bound of 260 GeV.

\subsubsection{Vector-like up quarks}
For model VI we use the results of \cite{CDFtquark} searching for pair produced $t' \, \bar{t}'$ that decay into excess $\ell E_T \! \! \! \! \! \! \! /  \, \, \, + jets$ events from $t' \rightarrow W q$ decays
at the Tevatron. This study selects for one and only one isolated muon or electron with $E_T$ or $P_T$ respectively greater than $25 \, {\rm GeV}$ as a trigger lepton.
The remaining trigger requires $E_T \! \! \! \! \! \! \! /  \, \,\, > 20 \, {\rm GeV}$  and at least four jets with  $E_T   > 20 \, {\rm GeV}$ and $|\eta| < 2.0$.
Again we rescale $\sigma(t' \, \bar{t}')$ given in \cite{CDFtquark} by $1/2$ as only the left-handed vector-like quarks decay into final states that contain $W^\pm$.

We also rescale to correct for the appropriate branching ratio in model VI to the triggered-on final states. In this model the effective branching ratio for one of the produced $V^u_L$
to give the signal ($\beta_t$) is given by
\bea
\beta_t &\simeq& {\rm BR}(V_L^j \rightarrow W^+ \, d_L^i) \times \left[ {\rm BR}(\bar{V}_L^j \rightarrow W^- \, \bar{d}_L^i) \times {\rm BR}( W^-  \rightarrow {\rm hadrons})\right] \nn \\
& \,& \, \, +   \, {\rm BR}(V_L^j \rightarrow W^+ \, d_L^i) \times \left[{\rm BR}(\bar{V}_L^j \rightarrow Z  \, \bar{u}_L^j) \times {\rm BR}( Z  \rightarrow {\rm hadrons}) \right] \nn \\
& \,& \, \, +   \, {\rm BR}(\bar{V}_L^j \rightarrow W^- \, \bar{d}_L^i) \times \left[{\rm BR}(V_L^j \rightarrow Z  \, u_L^j) \times {\rm BR}( Z  \rightarrow {\rm hadrons}) \right] \\
&\simeq& 0.61\ . \nn
\eea
Rescaling the NLO $\sigma(p \, \bar{p} \rightarrow t' \, \bar{t}')$ curve in Fig. 2 of \cite{CDFtquark}  by $\tfrac{N_f}{2}\beta_t$ we have conservatively obtained a 95\% CL lower bound for the vector-like up mass of 325 GeV
for model VI from the upper limit of the expected $95 \%$ CL region.

\subsection{Electroweak precision data constraints}\label{EWPD}

\subsubsection{EWPD fit}
EWPD constraints are weak as the singlet vector-like quarks do not directly break custodial symmetry and for models I, VI
contributions to the $\Pi^{\mu \nu}_{ab}$ with more than one mass scale in the loop are suppressed by $\epsilon_i$.
In considering EWPD we neglect NP effects that are suppressed by $\epsilon_i$
and neglect the effects of spurion insertions that break the flavour symmetry proportional to $\eta$.
For model VI we assume a mild hierarchy
$m_2 > m_3$ so that the top Yukawa $\epsilon_i$ can also be neglected.
With these assumptions the contribution to the vacuum polarization of the $Z$ (proportional to
$g^{\mu \nu}$) from a singlet vector-like quark of charge $Q$ is
\bea
\!\!\!\!\!\!\!\!\Pi^{\mu \, \nu} _{ZZ}(q^2)=  \frac{g_2^2 \sin^2\theta_W N_c N_f  \, Q^2 g^{\mu \, \nu}}{18 \, \pi^2} \left[6 (A_0(m_V^2) - m_V^2) + q^2 - 3 (2 m_V^2 + q^2) B_0(q^2, m_V^2 ,m_V^2)\right].
\eea

This result is given in terms of the Passarino--Veltman functions \cite{Passarino:1978jh} with standard definitions. All other vacuum polarizations can be determined in terms of this result
through taking appropriate derivatives and rescaling the couplings. We then construct the
EWPD parameters \cite{Holdom:1990tc,Peskin:1990zt,Golden:1990ig}. We use $\rm  \, STUVWX$  \cite{Burgess:1993mg,
Maksymyk:1993zm} as the constrained masses are expected to be in the 100 $\rm GeV$ range.
The results for models I, VI are reported in Table III.
\vspace{0.4cm}
\begin{table}[h]
\begin{center}
\begin{tabular}[t]{|c|c|c||c|c|c|}
  \hline
  \hline
     \multicolumn{6}{|c|}{$\rm STUVWX$ fit results} \\ \hline
    model & 68 $\%$ CL & 95 $\%$ CL & model & 68 $\%$ CL & 95 $\%$ CL \\  \hline
I &  \,  82 \, {\rm GeV} \,& \,  81 \, {\rm GeV} \, & VI &  \, 147 \, {\rm GeV}\, & \,102 \, {\rm GeV}\,\\ \hline
\hline
\end{tabular}
\end{center}
\caption{The results of an $\rm STUVWX$ fit as defined in \cite{Burgess:2009wm} where the CL regions are defined for six parameters
through the cumulative distribution function. The constraints on the vector-like up quarks are stronger as $Q = 2/3$.}
\end{table}

\subsubsection{$R_b$ constraints}

The oblique EWPD fit \cite{Burgess:2009wm} used in the previous section did not include deviations in the parameter $R_b  \equiv \Gamma(Z \rightarrow \bar{b} \, b)/\Gamma(Z \rightarrow {\rm hadrons})$.
Writing the coupling of the $Z$ boson to the quarks as
\bea
\mathcal{L}_Z = - \sqrt{g_1^2 + g_2^2} \, Z^\mu \, \bar{q} \, \gamma_{\mu} ( f_{L,q} \, P_L + f_{R,q} \, P_R) \, q\ ,
\eea
where $P_{L/R} = (1 \mp \gamma_5)/2$,  the tree level couplings to the $Z$ are given by
\bea
f_{L,q}^0 = T^3_q - \sin^2 \theta_W \, Q\ , \quad \quad f_{R,q}^0 =  - \sin^2 \theta_W \, Q\ .
\eea
Deviation in the left-handed coupling, due to mixing with the vector-like quarks, is given by $f_{L,b} = f_{L,b}^0 + \delta \, f_{L,b}$\,.
This leads  to a deviation in $R_b$ that can be approximated by \cite{Gresham:2007ri}
\bea
\delta \, R_b \simeq 2 \, R_b^0 (1 - R_b^0) \, \left(\frac{f_{L,b}^0 \, \delta f_{L,b}}{(f_{L,b}^0)^2 + (f_{R,b}^0)^2} \right) \simeq -0.78 \, \delta f_{L,b}\ .
\eea

For the vector-like down quarks the largest anomalous contribution is given by a tree level shift in the $Z \, \bar{b} \, b$ coupling.
In model I, a positive deviation $\delta f_{L,b} =\epsilon_3^2/2$ is predicted.
As the measured value of $R_b$ at the $Z$ pole and the SM predictions \cite{Amsler:2008zzb} are
\bea
R_b^{meas} = 0.21629 \pm 0.0066\ , \quad \quad R_b^{SM} = 0.21578 \pm 0.00010\ ,
\eea
the $1 \sigma$ bound on the deviation of $R_b$ from its experimental measurement due to the SM and this deviation $\delta f_{L,b}$
is given by $\, -1.5 \times 10^{-3} < \delta f_{L,b} < 2 \times 10^{-4}$. This translates into the following bound on the parameters in model I
\bea
\left(\frac{m_3^d}{m_2^d}\right)^2 \, \left(\frac{m_b^2}{v^2} \right)   < 2 \times 10^{-4}\ .
\eea

For the vector-like up quarks, determining the contributions to $R_b$
involves modifications of the SM diagrams for the top contributions to $R_b$ and new diagrams where $Z \rightarrow \bar{t} \, V_L^3$ in the loop.
The required calculation has been performed in sufficient generality
before and we use the results of
\cite{Bamert:1996px} for the large $m_t^2 ,m_V^2 \gg m_W^2$ limit.
We retain only the effects due to $(V^u_L)^3$ that are proportional to $m_t^2/v^2$.
Recalling that for model VI the third generation singlet quark has a mass $m_V^2 = (m^u_2)^2 \, (1 + \epsilon^2_3)$, the contribution to $R_b$ is given by
\bea
 \delta f_{L,b} &=& \frac{\alpha}{16 \, \pi \, \sin^2 \theta_W} \, \left(\frac{\sqrt{2} \, m^u_3 \, m_t}{m^u_2 \, v}\right)^2 \, \left[ \frac{3 (m_V^2 - m_t^2) + 2 \, m_V^2 \, m_t^2/m_W^2}{m_V^2 - m_t^2} \, \log \left(\frac{m_V^2}{m_t^2}\right)  - 2 \frac{m_t^2}{m_W^2}\right].
\eea
\\

\subsection{Flavour constraints}\label{flavour}
MFV forbids FCNC contributions to meson mixing at leading order in the MFV expansion that have been discussed in the literature \cite{AguilarSaavedra:2002kr,Barger:1995dd}.
However, the effects of SM quarks mixing with virtual vector-like quarks and anomalous SM couplings can still contribute to deviations from the SM phenomenology
of meson mixing measurements. We first consider these constraints before going to $b \rightarrow s \, \gamma$ constraints.

\subsubsection{Vector-like down quarks: Meson mixing}
Consider the additional contribution to the effective Hamiltonian for
$|\delta S| = 2$, $K^0 - \bar{K}^0$ mixing, which can be characterized by an extra contribution to the dominant operator
\bea
\delta \mathcal{H}_{|\delta S| = 2} = (C_{SM} + C_V)  \, \left(\bar{d}_L \, \gamma^\nu \, s_L \right) \left(\bar{d}_L \, \gamma_\nu \, s_L \right) + h.c.\ .
\eea

As the virtual quarks are charge $+ 2/3$, the potentially largest flavour changing effects for $K^0 - \bar{K}^0$ mixing in model I (when $\eta \ll 1$ and top Yukawa breaking of the flavour symmetry is neglected) come about through the non-unitarity of the effective CKM matrix.
The anomalous Wilson coefficient (following the notation of  \cite{AguilarSaavedra:2002kr}) is given by
\bea
C_V =  - \frac{G_F}{\sqrt{2}} \, \frac{\alpha}{4 \, \pi \, \sin \theta_W} \left[8 \, \sum_{\alpha = c,t} \lambda_\alpha \,  B_0(x_\alpha)  \, X_{sd} + X_{sd}^2 \right],
\eea
where the effective Inami-Lim function \cite{Inami:1980fz,Buchalla:1995vs} and the non-unitarity of $V_{CKM}$ is given by
\bea
B_0(x_\alpha) = \frac{1}{4} \left[ \frac{x_\alpha}{1 - x_\alpha} + \frac{x_\alpha}{(x_\alpha - 1)^2} \, \log(x_\alpha) \right], \quad \quad X_{sd} =  \left(\frac{m_3^d}{m_2^d}\right)^2 \sum_{\alpha = u,c,t} \lambda_{sd}^\alpha  \, \left(\frac{m_s^2 + m_d^2}{v^2}\right),
\eea
and we have defined $x_\alpha = (m^d_\alpha)^2/m_W^2$, $ \lambda^\alpha_{sd} = V^\star_{\alpha \, s} \, V_{\alpha \, d}$.
Note that the anomalous Wilson coefficient is SM CKM and Yukawa suppressed
as expected due to MFV. We observe that in model I non-unitarity effects  of this form on $K^0, \, B^0_s, \, B^0$ mixing vanish as
the deviation from unitarity is proportional to off-diagonal elements of the (decoupled) square of the SM CKM matrix $ \sum_{\alpha = u,c,t} \lambda_{sd}^\alpha$.
This effect is easily seen in a simple example
with a unitary matrix $P$ and a non-unitary matrix $Q$ where
 \begin{align}
P=\left(
    \begin{array}{cc}
      A & B \\
      C & D\\
    \end{array}
  \right)\ , \ \ \ Q =\left(
    \begin{array}{cc}
     X & 0 \\
      0 & Y \\
    \end{array}
  \right)\ ,
\end{align}
then
\vspace{-0.4cm}
 \begin{align}
(P + P \, Q)^\dagger \, (P + P \, Q) =  \left(
    \begin{array}{cc}
      1+ 2 \, X + X^2 & 0 \\
      0 & 1+ 2 \, Y + Y^2\\
    \end{array}
  \right)\ .
\end{align}
The deviation from unitarity in the square of the effective CKM matrix is only in the diagonal entries and does not effect $K^0, \, B^0_s, \, B^0$ mixing through
terms proportional to $X_{sd}$.
This form of non-unitarity has been termed pseudo-orthogonality in some past studies of vector-like down quarks \cite{Bjorken:2002vt} from $\rm E_6$ compactifications.
We adopt this nomenclature for this property of the effective CKM matrix.

Measurements of $D^0$ oscillations, however, proceed through charge $-1/3$ quarks and are sensitive to virtual contributions from the singlet
quarks in model I. The Lagrangian term of interest is
\bea
\mathcal{L} = g_2 \, \left(\frac{m_j \, m^d_3}{v \, m^d_2} \right) \, \bar{u}_L^i \, (V_{CKM})_i^j  \, W^+ \! \! \! \! \! \! \!\!\! / \, \, \,  \, \, \,  (V_L)_j  + h.c.\ ,
\eea
where the index on $m_j$ is again a label coincident with the flavour index and is not contracted.
The above CKM entries correspond to the SM entries as the vector-like quarks decouple in the limit $m^d_2 \rightarrow \infty$.
These CKM entries are equal to the measured CKM values up to consistently neglected higher order $(m^d_j)^2/v^2$ terms in model I.
Considering CKM and Yukawa suppression, the vector-like quark contribution is dominated by the contributions from the third generation
vector-like quarks and the corresponding contribution to $x_D$ (we use the definition of this mixing parameter given in \cite{Golowich:2007ka}) is
\bea
x_D^V \simeq \frac{2 \, G_F^2 \, m_W^2 \, f_D^2 \, M_D}{3 \, \pi^2 \, \Gamma_D} \, B_D \, (V_{cb}^\star \, V_{ub})^2 \, r_1(m_c, m_W) \, \left(\frac{m_b \, m^d_3}{v \, m^d_2} \right)^4 f(x_V)\ ,
\eea
\noindent{where $x_V = m_V^2/m_W^2$ and $f(x_V) \rightarrow x_V \, [1 + 6 \log(x_V)]$ in the large $x_V$ limit, which is applicable considering collider constraints
discussed in Section \ref{collider}.} We use the values of the CLEO-c determination \cite{Artuso:2005ym} of $f_D = 222.6 \pm 16.7^{+2.3}_{-2.4} \, {\ \rm MeV}$ and the lattice calculation of
$B_D = 0.83$ reported in \cite{Gupta:1996yt}. The renormalization group running   for a LO calculation  is given by \cite{Ciuchini:1997bw}
\bea
r_1(\mu,M) = \left(\frac{\alpha_s(M)}{\alpha_s(m_t)} \right)^{2/7} \left(\frac{\alpha_s(m_t)}{\alpha_s(m_b)}  \right)^{6/23} \left(\frac{\alpha_s(m_b)}{\alpha_s(\mu)}  \right)^{6/25}.
\eea
The remaining parameters we take from the reported results in the PDG. The measured value of  $x_D$ from the Belle collaboration in the
analysis of $D^0 \rightarrow K_S \, \pi^+ \, \pi^-$  \cite{Abe:2007rd} is given by
\vspace{-1mm}
\bea
x_D = (0.80 \pm 0.29 \pm 0.17) \times 10^{-2}\ ,
\eea
while fits to the HFAG database (without this measurement as a prior) give $x_D = 8.4^{+3.2}_{-3.4} \, \times 10^{-3}$ \cite{Golowich:2007ka} from the short distance SM OPE contribution.
Long distance contributions have been estimated to be of the order of $10^{-3}$ \cite{Donoghue:1985hh} but are difficult to reliably calculate. The contribution
of the vector-like down quarks to this quantity is far too small to be detected due to Yukawa and CKM suppression.
\vspace{-2mm}
\subsubsection{Vector-like up quarks: Meson mixing}

The effect of charge $Q = 2/3$  vector-like up quarks on $D^0$ mixing vanishes due to the pseudo-orthogonality nature of the modified CKM matrix.
However, these vector-like quarks can contribute to $K^0,B^0,B^0_s$ meson mixing through virtual contributions in box diagrams.
We ignore $\eta$ corrections to the Higgs coupling in these diagrams as they represent small Yukawa suppressed symmetry breaking terms.
We only retain the contributions from $(V^u_L)^3$  whose effects are proportional to $m_t^2/v^2$.
Using the results of \cite{AguilarSaavedra:2002kr} the new contributions to $M_{12}^K$  for $K^0 - \bar{K}^0$ mixing are given by
\bea
M_{12}^K = \frac{G_F^2 \, m_W^2 \, f_K^2 \, \hat{B}_K \, m_{K}}{12 \, \pi^2} \, \left[ \lambda_V^2 \, \eta_{V} \, S_0(x_V) + 2 \, \lambda_c \, \lambda_V \, \eta_{cV} \, S_0(x_c,x_V)  + 2 \, \lambda_t \, \lambda_V \, \eta_{t,V} \, S_0(x_t, x_V) \right]. \nn
\eea
The functions $S_0(x),S_0(x,y)$ are the usual
Inami-Lim functions given by \cite{Inami:1980fz,Buchalla:1995vs}
\bea
S_0(x) &=& \frac{4 \, x - 11 \, x^2 + x^3}{4(1-x)^2} - \frac{3 x^3}{2(1-x)^3} \, \log{x}\ , \nn \\
S_0(x,y) &=& - \frac{3 \, x \, y}{4(x-1) (y-1)} + \frac{x \, y \ (x^2 - 8 x + 4)}{4 \, (x-1)^2 \, (x-y)} \,\log{x} + \frac{x \, y \, ( y^2 - 8 y + 4)}{4(y-1)^2 (y-x)} \, \log{y}\ .
\eea
The contribution of this expression to $\Delta \epsilon_K$ is
\bea
|\Delta \, \epsilon_K| = C_{\epsilon}  \, \hat{B}_K \, \epsilon_3 \, {\rm Im} \left[ \epsilon_3 \, \lambda_t^2 \, \eta_{V} \, S_0(x_V) + 2 \, \lambda_c \, \lambda_t \, \eta_{cV} \, S_0(x_c,x_V)  + 2 \, \lambda_t^2  \eta_{t,V} \, S_0(x_t, x_V) \right].
\eea
Recall $\epsilon_3 = \frac{\sqrt{2} \, m_t \, m^u_3}{v \, m^u_2}$.
We take NLO values for the approximated QCD running from the top mass which gives $\eta_{V} = 0.58$, $ \eta_{cV} \sim \eta_{ct} \simeq 0.47$ and $ \eta_{tV} \sim \eta_{tt} \simeq 0.57$
from \cite{AguilarSaavedra:2002kr,Herrlich:1996vf}.
Using the measured values $m_K = 497.6 \,
{\rm MeV}$, $f_K = (156.1 \pm 0.8) {\ \rm MeV}$, $(\Delta M_K)_{exp}
=  (3.483 \pm 0.006) \times10^{-12} {\ \rm MeV}$ one obtains
\bea
C_{\epsilon} = \frac{G_F^2 \, F_K^2 \, m_K \, M_W^2}{6 \, \sqrt{2}
\, \pi^2 \, \Delta M_K} = 3.65 \times 10^4.
\eea
Further, lattice QCD \cite{Lubicz:2008am} gives the input $B_K(2
\, {\rm GeV}) = 0.54 \pm 0.05$. Considering that current measurements find $|\epsilon_K|_{exp} = (2.229 \pm 0.010)\times 10^{-3}$
while recent CKM fits lead to the SM prediction $|\epsilon_K|_{theory} = (1.8 \, \pm 0.5)  \times 10^{-3}$,
we find the $1 \sigma$ bound $~-0.07  \times 10^{-3}<  |\Delta \, \epsilon_K| < 0.93 \times 10^{-3}$.

We have also determined the constraints in model VI from $B^0 - \bar{B}^0$ mixing and $B_s^0 - \bar{B}_s^0$ mixing
from the results of \cite{AguilarSaavedra:2002kr}. We find that these constraints are negligible when compared to
kaon mixing and the other constraints discussed.

\subsubsection{$b \rightarrow s \, \gamma$ constraints}

Recall that the effective Hamiltonian (neglecting the light quark masses) in the SM is given by \cite{Inami:1980fz, Grinstein:1987vj}
\bea
\mathcal{H}_{\rm eff} &=&  \frac{2 \, G_F}{\sqrt{2}} \, V_{tb} \, V_{ts}^\star \, \sum_{i=1}^8 \, C_i(\mu) \, O_i \nn \\
&=& \frac{2 \, G_f}{\sqrt{2}} \, V_{tb} \, V_{ts}^\star \, A(x_t) \, \left[\frac{e \, m_b}{16 \, \pi^2} \right] \, \bar{s}_L \, \sigma^{\mu \, \nu} \, b_R \, F_{\mu \, \nu} + \cdots\ ,
\eea
where the Inami-Lim function for the operator $O_7$ above is
\bea
A(x) &=& -\frac{x}{3} \, \left[- \frac{1}{2} \, \frac{1}{(x-1)} + \frac{3}{2} \, \frac{1}{(x-1)^2} + \frac{3}{(x-1)^3} \right]  + \frac{x}{2} \left[\frac{1}{(x-1)} + \frac{9}{2} \, \frac{1}{(x-1)^2} + \frac{3}{(x-1)^3} \right] \nn \\
&\,& + \frac{x^2}{(x-1)^4} \, \log{x} - \frac{3}{2} \, \frac{x^3}{(x-1)^4} \, \log{x}
\eea
and $x_i = m_i^2/m_W^2$. For new contributions to $b \rightarrow s \, \gamma$ in linear MFV, the largest effect of the vector-like down quarks of
model I is the modification of the SM $\,W^\pm$ coupling, while for vector-like up quarks the largest modification
comes about through the third generation virtual vector-like quark in the usual loop contribution to  $O_7$.
In both models the largest contribution comes from the third generation vector-like quark.

For model I the largest new contribution to $O_7$ is
\bea
C_V^d = - \frac{4 \, G_f}{\sqrt{2}} \, V_{tb} \, V_{ts}^\star \, A(x_t) \,\left(\frac{m_b \, m^d_3}{v \, m^d_2} \right)^2,
\eea
while for model VI the largest new contribution is
\bea
C_V^u = \frac{4 \, G_f}{\sqrt{2}} \, V_{tb} \, V_{ts}^\star \, A(x_V) \,\left(\frac{m_t \, m^u_3}{v \, m^u_2} \right)^2.
\eea
One can use the results of \cite{Grzadkowski:2008mf} for the contribution of such a Wilson coefficient
to ${\rm BR}(\bar{B} \rightarrow X_s \, \gamma)_{E_\gamma > 1.6 \, {\rm GeV}}$ given by
\bea
{\rm BR}(\bar{B} \rightarrow X_s \, \gamma)_{E_\gamma > 1.6 \, {\rm GeV}} = 3.15 \pm 0.23 - 8.0 \, \left(\frac{\sqrt{2} \, C_V^{u,d}}{4 \, G_f \, V_{tb} \, V_{ts}^\star}\right).
\eea
Comparing to the current world experimental average \cite{Barberio:2007cr}
\bea
{\rm BR}(\bar{B} \rightarrow X_s \, \gamma)_{E_\gamma > 1.6 \, {\rm GeV}} = 3.55 \pm 0.24^{+0.09}_{-0.10} \pm 0.03\ ,
\eea
we obtain a $1 \sigma$ bound of $\ -0.17 < \left(\frac{\sqrt{2} \, C_V^{u,d}}{4 \, G_f \, V_{tb} \, V_{ts}^\star}\right) < 0.07$.
\\

\subsection{Combined constraints}\label{combo}

The constraints on the model parameters $m_2$ and $m_3$ that we have derived are presented in Fig. 1.
\vspace{5mm}
\begin{figure}[hbtp]
\centerline{\scalebox{0.9}{\includegraphics{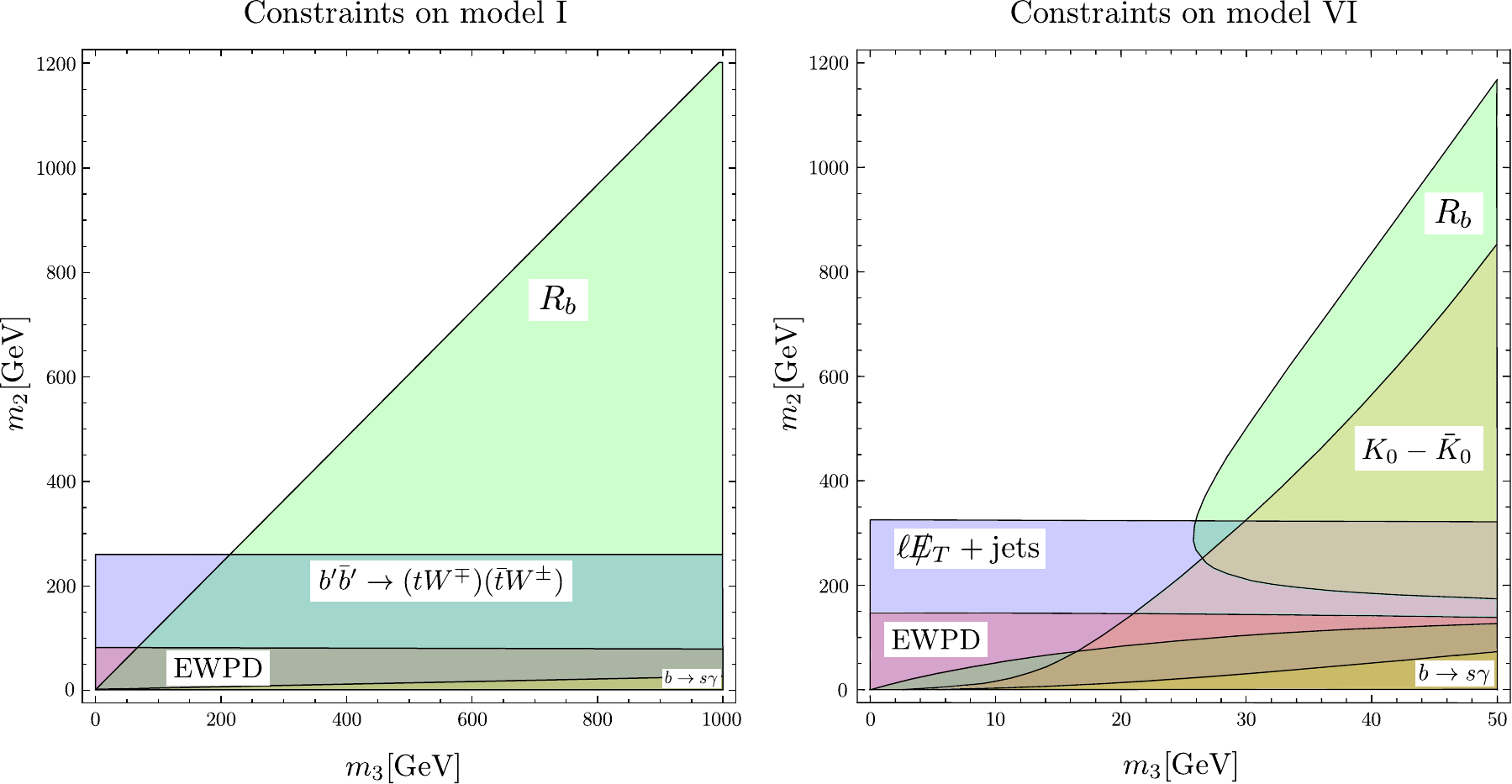}}}
\caption{The constraints on model I (left) and model VI (right). The shaded regions are excluded by the various labeled observables
considering a $1 \sigma$ deviation in the labeled measurement adding the SM and the vector-like quark contribution; for EWPD
a $68 \%$ CL exclusion region is shown, while for Tevatron constraints the $95 \, \%$ CL exclusion region is shown.}
\end{figure}
Shown are the $68 \%$ CL regions from the EWPD fit, the $95 \% $ CL
constraints from direct searches at the Tevatron utilizing the  $b'  \, \bar{b'}  \rightarrow (t W^{\mp}) \, (\bar{t} W^{\pm})$ decays for
model I and  $\ell E_T \! \! \! \! \! \! \! /  \, \, \, + {\rm jets}$ constraints from $t' \rightarrow W q$ decays for model VI.
Also shown is the $1 \sigma$ bound for $K_0 - \bar{K}_0 $ mixing, $R_b$ and $b \rightarrow s \,  \gamma$.
Note that the parameter space shown for model VI is characterized by $m_2 >m_3$ which is consistent with the assumed mild hierarchy
allowing the vector-like quark Lagrangian construction without retaining all orders in $\epsilon_3$.
Fig. 2 shows how the flavour and $R_b$ constraints  are relaxed for models I and VI when larger deviations are allowed by comparing the
$1\sigma$ and $2\sigma$ regions.
\vspace{5mm}
\begin{figure}[hb]
\centerline{\scalebox{0.9}{\includegraphics{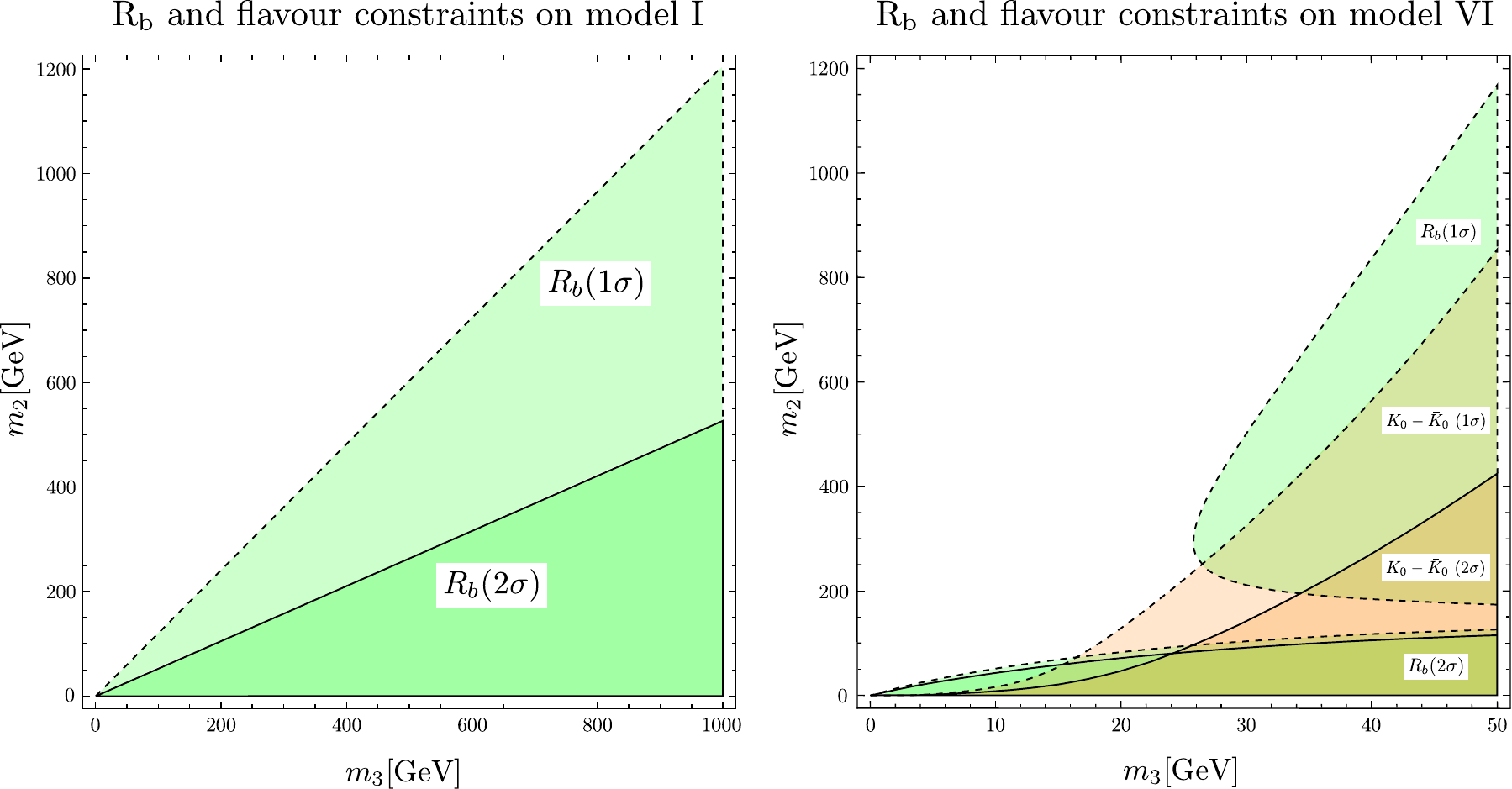}}}
\caption{Comparison of the $1\sigma$ (dashed line) and $2\sigma$ (solid line) meson mixing and $R_b$ constraints for models I (left) and model VI (right). }
\end{figure}

For comparison, to illustrate the relative flavour unnaturalness of the remaining models consider models III and VIII, whose mixing angles and mass spectrum are given in the Appendix.
(We compare these models simply for ease of comparison as only two Lagrangian parameters are present for these
models.)
Models III and VIII  do not satisfy our flavour naturalness criteria.
One has to  consider Lagrangian parameters $\sqrt{m_2^2 + m_3^2 } \gtrsim 10^6 \, {\rm GeV}$
for both models so that the lightest vector-like quark is $ \gtrsim 100 \, {\rm GeV}$ considering direct production bounds.
The vector-like quark  masses would then be in a pattern $m_1^d : m_2^d : m_3^d$ given by $ \sim 10^2 : 10^3 : 10^5 \, {\rm GeV}$
in model III; in model VIII the pattern would be similar with the heaviest vector-like quark being $\sim  \, 10^6 \, {\rm GeV}$.
Only the phenomenology of the lightest vector-like quark states would be readily accessible at LHC or the Tevatron.

It is convenient to change to a polar coordinate system when determining the constraints on this model, where $r \sim \sqrt{m_2^2 + m_3^2}$ and $m_3 = m_2 \, \tan{\phi}$.
Viable parameter space requires $r \gtrsim  10^6  \, {\rm GeV}$ and $\phi \ll 1$, where $m_V^i \sim \sqrt{m_0^i \, r}$. Expanding in large $r$ and small $\phi$ one finds
 \begin{align}
U=\left(
    \begin{array}{cc}
       1 - \left(\frac{r^2 + v^2}{2 \, r^2}\right) \phi^2 & \ \ -  \left(\frac{2 \, r^2 + v^2}{2 \, r^2}\right)\phi\\
      \left(\frac{2 \, r^2 + v^2}{2 \, r^2}\right)\phi & \ \ 1 - \left(\frac{r^2 + v^2}{2 \, r^2}\right) \phi^2  \\
    \end{array} \right)\ .
\end{align}
Using this parametrization the constraint from $R_b$ for model III is given by
\bea
\phi^2 \, \left(\frac{r^2 + v^2}{2 r^2} \right) < 1.5 \, \times 10^{-3},
\eea
which gives $\phi < 0.055$, while the direct production bound at the Tevatron is given by $r > 7.3 \times 10^6 \, {\rm GeV}$.
The Lagrangian parameters must be chosen consistently with these strong constraints for this model to be viable and any
LHC vector quark signal in early runs would have to correspond to this small subset of the parameter space.

For model VIII the correction to $f_{L,b}$ is
\bea
 \delta f_{L,b} &=& \frac{\alpha}{16 \, \pi \, \sin^2 \theta_W}  \, \left(\frac{m_V^2 - m_t^2}{m_W^2} + 3 \, \log \left[\frac{m_V^2}{m_t^2} \right] \right)  \nn \\
 &+& \frac{\alpha \, \phi  \, \left( 3 \, m_t^4  - 4 m_V^2 \, m_t^2 + m_V^4\right)}{16 \, \pi \, \sin^2 \theta_W \, (m_V^2 - m_t^2)\, m_W^2}   +  \frac{\alpha \, \phi  \, \left[ (m_V^2  - 3 m_W^2) \, m_t^2 + 3 m_V^2 \, m_W^2 \right]}{8\, \pi \, \sin^2 \theta_W \, (m_V^2 - m_t^2)\, m_W^2}  \, \log \left[\frac{m_V^2}{m_t^2} \right]\ .
 \eea
The vector-like up quark collider constraint from $V_L \rightarrow W \, q_L$ decays \cite{CDFtquark} directly applies for the lightest $Q = 2/3$ vector-like quark giving $m_V > 325 \, {\rm GeV}$.
This translates into a constraint  $r >2 \times 10^7 \, {\rm GeV}$. The allowed values of $\phi$ are $\phi <  2 \times 10^{-2}$ for $m_V = 325 \, {\rm GeV}$  which monotonically decreases to $\phi <  2 \times 10^{-3}$ for $m_V = 1200 \, {\rm GeV}$. Again, the Lagrangian parameters must be chosen consistently with these strong constraints for this model to be viable.

\section{Production at LHC}

The discovery of vector-like quarks at LHC has been studied extensively in the literature \cite{Andre:2003wc,Rosner:1985hx,Barger:1985nq,AguilarSaavedra:2005pv,Grossman:2007bd,AguilarSaavedra:2009es,Berger:2009qy}.
Generally, the discovery signatures rely on the decay of the vector-like quarks producing a $W$ or
$Z$ boson  giving leptonic tags as well as SM quarks that initiate jets. Studies at LHC are likely to follow the Tevatron studies
of this form \cite{Aaltonen:2007je,CDFbquark,CDFtquark}
but also have the opportunity to apply new theoretical approaches such as employing jet mass \cite{Skiba:2007fw,Holdom:2007nw}

The results of these various studies are applicable to
the MFV models studied in this paper with the appropriate rescaling of the decay widths in terms of the known
mixing parameters. However, as most of these studies have employed simulations of LO QCD, in this section we present the production
cross sections for the allowed remaining parameter space in Fig. 1 using the analytic inclusive NLO QCD production results of \cite{Czakon:2008ii} which include the partonic
production channels
\bea
q + \bar{q} \rightarrow \bar{V}_{L/R} \,  V_{L/R} + X\ , \nn \\
g + g \rightarrow \bar{V}_{L/R} \,  V_{L/R} + X\ , \nn \\
g + q \rightarrow \bar{V}_{L/R} \,  V_{L/R} + X\ .
\eea
We correct this result with a factor of $N_f/2$ because of the three flavours of the vector-like quarks and the chiral factor, as only $V_L$ will typically
give gauge bosons that can be triggered-on (see Section \ref{collider} for details). The effective cross section is
\bea
\sigma_{\rm eff} = \frac{N_f}{2} \, \left(\sigma_{q\,\bar{q}} + \sigma_{gg} + \sigma_{gq} \right).
\eea
We show the sum of these inclusive cross sections in Fig. 3 (left) determined using the NNLO PDF's of MSTW  \cite{Martin:2009iq} and
four-loop running $\beta$ function results of \cite{vanRitbergen:1997va} to run from the reference value $\alpha_s(M_Z) = 0.1135$ \cite{Abbate:2010vw}.
We vary the renormalization scale of the evaluation between $m_V/2 < \mu < 2 \, m_V$ to define a scale dependent error on the
production cross section. The results are shown in Fig. 3 (left) and the effective $\sigma$ is presented in Table IV for $\sqrt{s} = 7,10,14 \, {\rm TeV}$.
We additionally show the various contributions of the partonic production channels to the inclusive pair production effective cross section
for $\sqrt{s} = 10 \, {\rm TeV}$
also in Fig. 3 (right).
\vspace{3mm}
\begin{figure}[htbp]
\centerline{\scalebox{0.8}{\includegraphics{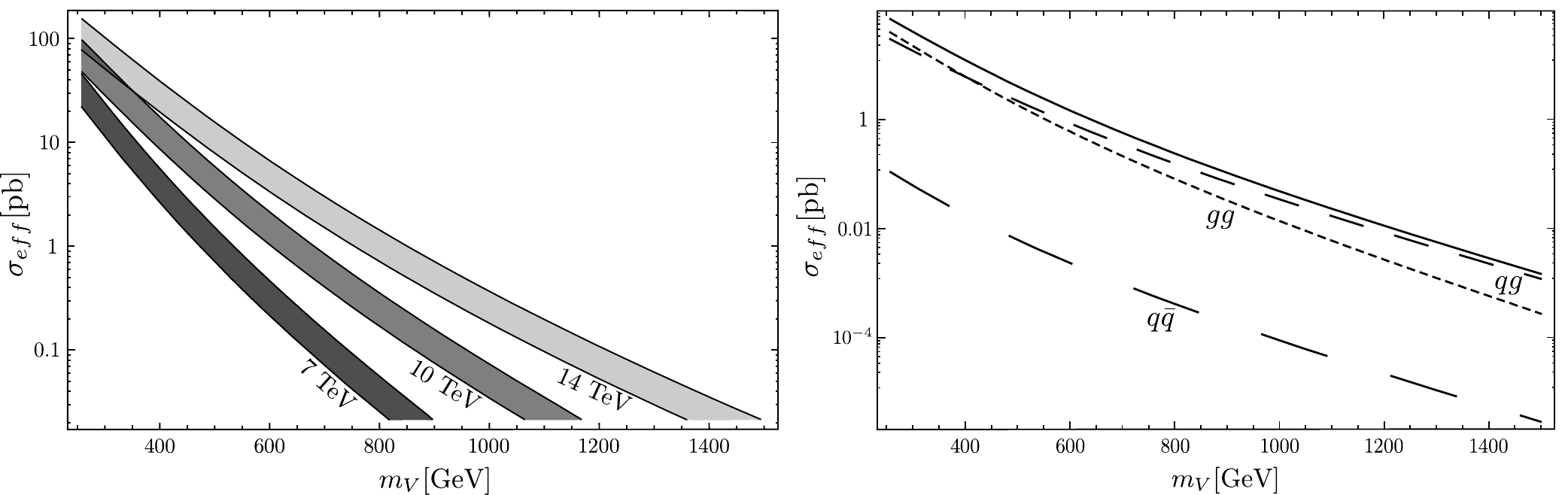}}}
\caption{Left figure: The effective NLO QCD pair production cross sections for $\sqrt{s} = 7, \, 10, \, 14 \, {\rm TeV}$, where the error band is
defined through the variation  of the renormalization scale $m_V/2 < \mu < 2 \,m_V$. Right figure: The various contributions to the effective NLO QCD pair production cross sections for $\sqrt{s} = 10 \, {\rm TeV}$. The solid line is the sum of all contributions.
The short dashed line is inclusive pair production through $g\, g \rightarrow V_{L} \, \bar{V}_L$, while the long dashed line is inclusive pair production through $q\, \bar{q} \rightarrow V_{L} \, \bar{V}_L$,
which is highly suppressed due to the antiquark PDF suppression. Production through $q \, g \rightarrow V_{L} \, \bar{V}_L$ dominates for high masses.\\
}
\end{figure}
\begin{table}[htbp]
\begin{center}
\begin{tabular}[t]{|c|c|c|c|}
  \hline
  \hline
    & \multicolumn{3}{|c|}{$\sigma_{\rm eff}^{NLO}$} \\ \hline
    $m_V \, [{\rm GeV}]$ & $\sqrt{s} = 7 \, {\rm TeV}$  & $\sqrt{s} = 10 \, {\rm TeV}$ &  $\sqrt{s} = 14 \, {\rm TeV}$  \\  \hline
332 & $10^{+5}_{-3} \,{\rm \, pb} $ &  $27^{+12}_{-8} \,{\rm \, pb} $ & $53^{+23}_{-15} \,{\rm \, pb} $ \\ \hline
502 & $1.0^{+0.5}_{-0.3} \,{\rm \, pb} $ &  $4.0^{+1.8}_{-1.2} \,{\rm \, pb} $ & $11^{+5}_{-3} \,{\rm \, pb} $ \\ \hline
704 & $0.10^{+0.05}_{-0.03} \,{\rm \, pb} $ &  $0.56^{+0.26}_{-0.17} \,{\rm \, pb} $ & $2.0^{+0.9}_{-0.6} \,{\rm \, pb} $ \\ \hline
934 & $10^{+5}_{-3} \,{\rm \, fb} $ &  $82^{+39}_{-25} \,{\rm \, fb} $ & $0.39^{+0.17}_{-0.11} \,{\rm \, pb} $ \\ \hline
1181 & $1.0^{+0.6}_{-0.3} \,{\rm \, fb} $ &  $13^{+7}_{-4} \,{\rm \, fb} $ & $84^{+38}_{-24} \,{\rm \, fb} $ \\ \hline
\hline
\end{tabular}
\end{center}
\vspace{-0.3cm}
\caption{$\sigma_{\rm eff}$ for models I, VI from inclusive NLO QCD production. The error is the scale variation error. We use the NNLO PDF's of MSTW (see Section IV for details).}
\end{table}

A large amount of parameter space remains in each model that can be probed at LHC. The allowed regions of the model parameters
permit large $\sim \rm pb$ production cross sections, and thus significant early LHC event rates that are not ruled out by current Tevatron searches or flavour constraints. This is illustrated in Fig. 4 for $\sqrt{s} = 7 \, {\rm TeV}$ with the errors for the $\sqrt{s} = 7 \, {\rm TeV}$ contours and values of the effective cross section for
$\sqrt{s} = 10, \, 14 \, {\rm TeV}$ given in Table IV.
\vspace{0.1cm}
\begin{figure}[htbp]
\centerline{\scalebox{0.8}{\includegraphics{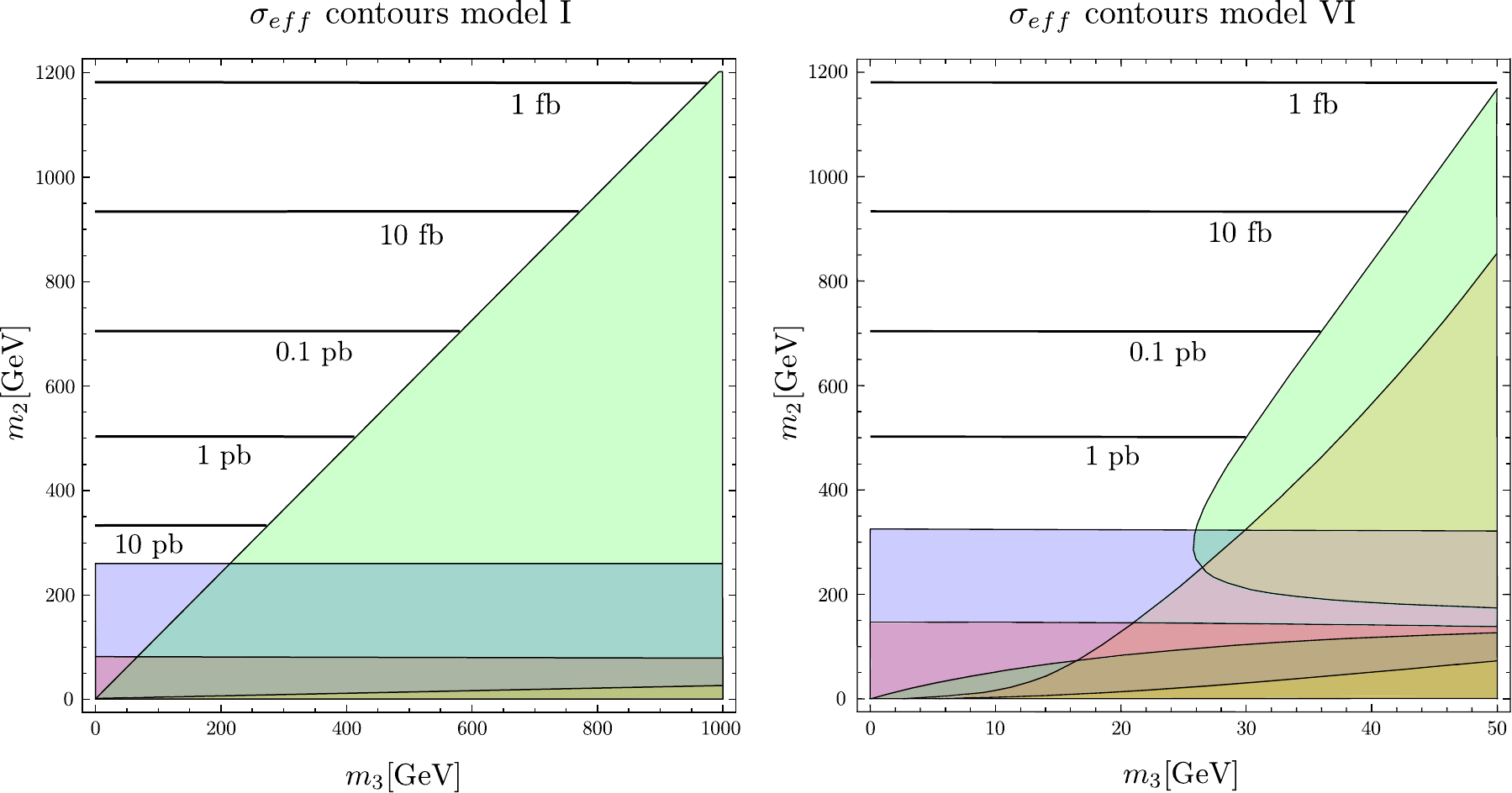}}}
\vspace{-0.2cm}
\caption{Contours of constant ${n_f} \sigma$ determined from analytic NLO QCD pair production in models I, VI as described in the text. The labels on the plot correspond to $\sqrt{s} = 7 \, {\rm TeV}$. See Table IV for the errors and
values for $\sqrt{s} = 7,10,14 \,  {\rm TeV}$ for the shown contours of constant $\sigma_{\rm eff}$. In model I the $\epsilon_i$ mass correction is negligible for all three generations.
We include in the contour the $\epsilon_3$ correction to the vector-like quark mass dependence for model VI. For the first two generation vector-like quarks in model VI the contours are coincident with
the left edge of the plotted contour and straight lines as the  $\epsilon_{1,2}$ mass correction is negligible.}
\end{figure}

\section{Conclusions}

We have examined all $\rm SU(2)_L$ singlet vector-like quark models with $Q = -1/3$ or $Q = 2/3$ that satisfy MFV while the spinor components are restricted to transform as a triplet
under one of the flavour groups of $ G_F$. We have directly solved these models and determined the mass spectra and mixing angles with the SM quarks.
This has identified two of these models as naturally phenomenologically viable due to the predicted mixing angles of the model and the effective SM CKM matrix structure.
These models are more phenomenologically viable
than the remaining models due to their predicted mixing which depends on their group structure and representation content.
However, we note that our analysis is done under the assumption that a perturbative analysis is appropriate
and relies on the mixing that these models experience with the SM quarks at leading order in the MFV expansion.
If $\eta \gtrsim 1$ so that flavour breaking insertions of the top Yukawa are not suppressed then a further analysis is required
to determine the constraints on the mass scale of these models and the LHC discovery potential.

The constraints on the two models were explored in detail using EWPD, Tevatron constraints and flavour observables and much parameter space
remains for LHC to explore. We have demonstrated this with the effective NLO QCD production cross sections for the viable parameter space.

Although the matter content we have studied has not been introduced to directly solve the hierarchy problem, such matter content
may be a component of a new physics sector that does solve the hierarchy problem. The earliest evidence
of such a sector at LHC could emerge from electroweak scale coloured states that satisfy known flavour constraints on the weak scale.

We emphasize  that if vector-like quarks are discovered in the early LHC era, they will likely have $m_V \lesssim 1 \, {\rm TeV}$ due to
event rate limitations. In such a scenario the compatibility of the vector-like quarks with the constraints we have explored
will be an important consistency check on a vector-like quark model consistent with an LHC signature. Due to their flavour structure, including the constraints of MFV, the models we have
explored and constrained in detail are particularly promising $\rm SU(2)_L$ singlet vector-like quark models for LHC discovery.

\subsection*{Acknowledgment}
We thank Mark Wise for inspiration and comments on the manuscript. M.T. thanks Maxim Pospelov for useful discussions.
Research at the Perimeter Institute is supported in part by the Government of Canada through NSERC and by the Provice of Ontario through MEDT.

\newpage

\appendix
\section{Remaining MFV model spectra and mixing angles}
\subsection*{Model II}
\begin{center}
\begin{tabular}{|c|}
  \hline
 $\kappa_1^d = g_d,\kappa_2^d  = {\bf 1},\kappa_3^d = {\bf 1}$\ ,\ \ \ $\mathcal{U}(V,L) = \mathcal{U}(V,R) = \mathcal{U}(d,L)$ \\ \hline
 $~~U = \frac{1}{\sqrt{\xi_2^2 + \xi_3^2}} \, \left(\begin{array}{cc}
\xi_2  &  - \xi_3 \\
\xi_3  & \xi_2 \\
\end{array}
\right)\ ,\ \ \
W = \left(\begin{array}{cc}
{\bf 1} & - \epsilon_i S \\
\epsilon_i \, S & {\bf 1}  \\
\end{array}
\right)\ ,$ \\
  $D^2  =  \left(
\begin{array}{cc}
(\mathcal{M}^{\rm phys}_d)^2 & 0 \\
0 &  (m^d_2)^2 + (m^d_3)^2  + (\mathcal{M}^{\rm phys}_d)^2 \, \left[\frac{ 2 \, \xi_1^2 \, \xi_2^2 +\xi_3^2+2\sqrt{2}\xi_1\xi_2\xi_3 }{\xi_2^2 +2\xi_1^2\xi_3^2-2\sqrt{2}\xi_1\xi_2\xi_3} \right] \\
\end{array}
\right)$ \\
  \hline
\end{tabular}
\end{center}
where
\bea
S = \tfrac{\xi_2}{\sqrt{2} \, \xi_3} \, \tfrac{\xi_3 + \sqrt{2} \, \xi_1 \, \xi_2}{\xi_2^2 + \xi_3^2}\ , \ \ \xi_{1,2,3}=\tfrac{m^d_{1,2,3}}{v}\ ,\ \
\mathcal{M}^{\rm phys}_d \equiv \mathcal{M}^0_d \, \sqrt{\frac{\xi_2^2 + 2 \, \xi_1^2 \, \xi_3^2  - 2 \sqrt{2} \, \xi_1 \, \xi_2 \, \xi_3}{\xi_2^2 + \xi_3^2} }\ .
\eea
In this model
we identify the three eigenvalues directly proportional to $\mathcal{M}^0_d$ to be identical with the SM quark masses. The corresponding decay widths are obtained through rescaling the couplings of Eqs. (18)--(20).
In this model, the mixing with the SM quarks is not naturally small  unless one is in the decoupling limit $m_2 \rightarrow \infty$.
For this one and the following models in the Appendix,  the results for the vector-like up quark models are trivial to obtain following the procedure discussed in Section \ref{vups}.
\\
\subsection*{Model III}
\begin{center}
\begin{tabular}{|c|}
  \hline
  $\kappa_1^d = {\bf 0}, \kappa_2^d = g_d, \kappa_3^d  = g_d$\ ,\ \ \ $\mathcal{U}(V,L) = \mathcal{U}(d,L)\  ,\  \mathcal{U}(V,R) =\mathcal{U}(d,R)$ \\ \hline
 $~~U= \left(
    \begin{array}{cc}
     \frac{C_{++}}{4\xi_2\xi_3\sqrt{\left(\frac{C_{++}}{4\xi_2\xi_3}\right)^2+1}} & \frac{1}{\sqrt{\left(\frac{C_{++}}{4\xi_2\xi_3}\right)^2+1}}  \\
     \frac{C_{+-}}{4\xi_2\xi_3\sqrt{\left(\frac{C_{+-}}{4\xi_2\xi_3}\right)^2+1}}
     & \frac{1}{\sqrt{\left(\frac{C_{+-}}{4\xi_2\xi_3}\right)^2+1}} \\
    \end{array}
  \right)\ ,\ \ \
W= \left(
    \begin{array}{cc}
     \frac{C_{-+}}{2\sqrt{2}\xi_3\sqrt{\left(\frac{C_{-+}}{2\sqrt{2}\xi_3}\right)^2+1}} & \frac{1}{\sqrt{\left(\frac{C_{-+}}{2\sqrt{2}\xi_3}\right)^2+1}}  \\
     \frac{C_{--}}{2\sqrt{2}\xi_3\sqrt{\left(\frac{C_{--}}{2\sqrt{2}\xi_3}\right)^2+1}}
     & \frac{1}{\sqrt{\left(\frac{C_{--}}{2\sqrt{2}\xi_3}\right)^2+1}} \\
    \end{array}
  \right)\ ,$ \\
  $D^2  =  \left(
\begin{array}{cc}
\left(\mathcal{M}_d^0\right)^2 \left(\xi_2^2+\xi_3^2-\tfrac{1}{2}\delta+\tfrac{1}{2}\right) & 0 \\
0 &  \left(\mathcal{M}_d^0\right)^2 \left(\xi_2^2+\xi_3^2+\tfrac{1}{2}\delta+\tfrac{1}{2}\right) \\
\end{array}
\right)$ \\
  \hline
\end{tabular}
\end{center}
where
\bea
\delta = \sqrt{4\xi_2^4+(8\xi_3^2-4)\xi_2^2+(2\xi_3^2+1)^2}\ , \ \
C_{\pm\pm} = -2\xi_2^2\pm 2\xi_3^2\pm \delta +1\ .
\eea
We identify three of these masses with the SM down quark masses and three of them with the new vector-like down quark states.
We discuss the constraints on this model in Section \ref{combo}.
\\
\subsection*{Model IV}
\begin{center}
\begin{tabular}{|c|}
  \hline
 $\kappa_1^d = {\bf 1}, \kappa_2^d = g_d^\dagger, \kappa_3^d  = {\bf 1}$\ ,\ \ \ $\mathcal{U}(V,L) = \mathcal{U}(d,R)  ,\  \mathcal{U}(V,R) = \mathcal{U}(d,L)$ \\ \hline
  $D^2  =  \left(
\begin{array}{cc}
(m^d_3)^2+\mathcal{O}[(\mathcal{M}_d^0)^2] & 0 \\
0 &  (m^d_1)^2+\mathcal{O}[(\mathcal{M}_d^0)^2] \\
\end{array}
\right)$ \\
  \hline
\end{tabular}
\end{center}
We refrain from including the mixing angles for this model in the Appendix due to their
length and complication, although we have determined them using the described procedure. The key point is that they do not provide naturally small non SM contributions to
quark couplings to the $W$ and $Z$. Also, as noted earlier, this model predicts non-hierarchical down quark masses and
can only be viable as long as  one chooses $m^d_i \lesssim m_d$, where $i = 1 \,{\rm or} \, 3$.

\subsection*{Model V}
\begin{center}
\begin{tabular}{|c|}
  \hline
 $\kappa_1^d = g_d,  \kappa_2^d = g_u,\kappa_3^d  = g_u$\ \\ \hline
 $~~U = \frac{1}{\sqrt{\xi_2^2 + \xi_3^2}} \, \left(\begin{array}{cc}
\xi_2  &  - \xi_3 \\
\xi_3  & \xi_2 \\
\end{array}
\right)\ ,\ \ \
W = \left(\begin{array}{cc}
{\bf 1} & - \epsilon_i C_i \\
\epsilon_i \, C_i & {\bf 1}  \\
\end{array}
\right)\ ,$ \\
  $D^2  =  \left(
\begin{array}{cc}
(\mathcal{M}^{\rm phys}_d)^2 & 0 \\
0 & 2 \, (\xi_2^2 + \xi_3^2) \, (\mathcal{M}^{\rm phys}_u)^2  +  (\mathcal{M}^{\rm phys}_d)^2 \, \frac{\xi_3^2 + 2 \xi_1^2 \, \xi_2^2 + 2 \sqrt{2} \, \xi_1 \, \xi_2 \, \xi_3}{\xi_2^2 + 2 \xi_1^2 \, \xi_3^2 - 2 \sqrt{2} \, \xi_1 \, \xi_2 \, \xi_3} \\
\end{array}
\right)$ \\
  \hline
\end{tabular}
\end{center}
where
\bea
C_i = \tfrac{v}{\mathcal{M}_u^i} \, \tfrac{\xi_2}{2\xi_3} \,  \tfrac{\xi_3+\sqrt{2}\,\xi_1\, \xi_2 }{\xi_2^2+\xi_3^2}\ , \ \ \xi_{1,2,3}=\tfrac{m^d_{1,2,3}}{v}\ ,\ \
\mathcal{M}^{\rm phys}_d \equiv \mathcal{M}^0_d \, \sqrt{\frac{2 \, \xi_1^2 \, \xi_3^2 + \xi_2^2 - 2 \sqrt{2} \, \xi_1 \, \xi_2 \, \xi_3}{\xi_2^2 + \xi_3^2} }\ .
\eea
In this model it is not possible to choose $\mathcal{U}(V,L/R)$ in such a way that the $3 \times 3$ subblocks of Eq. (\ref{diag}) are diagonal. Nevertheless, we take $\,\mathcal{U}(V,L) = \mathcal{U}(d,L),\  \mathcal{U}(V,R) = \mathcal{U}(u,R)$ and write the two left submatrices in a diagonal form and the two right submatrices as diagonal matrices multiplied by $V_{CKM}$. We use the Wolfenstein parametrization and neglect all terms of order $\mathcal{O}(\lambda^2)$ and higher.
We assume that  $m_d^0,m_u^0$ are approximately equal to the SM physical masses up to small corrections,  expanding in $m_i^0$ as usual
to simplify the eigenvalues and eigenvectors.
The rotation matrices in this model are identical in form as in model II
with the replacement $S \rightarrow C_i$.
The physics of this model is substantially the same, and unnatural with respect to flavour constraints. The only difference is that the light up quark masses receive a large correction to their Higgs coupling as
$C_i \gg 1$ due to $\mathcal{M}_u^{1,2} \ll v$.
\\
\\



\begin{thebibliography}{99}




\bibitem{Georgi:1972bb}
  H.~Georgi and S.~L.~Glashow,
  Phys.\ Rev.\  D {\bf 6}, 429 (1972).
\vspace{-0.3mm}

\bibitem{Frampton:1999xi}
  P.~H.~Frampton, P.~Q.~Hung and M.~Sher,
  Phys.\ Rept.\  {\bf 330}, 263 (2000)
  [arXiv:hep-ph/9903387].
\vspace{-0.3mm}

\bibitem{AguilarSaavedra:2002kr}
  J.~A.~Aguilar-Saavedra,
  Phys.\ Rev.\  D {\bf 67}, 035003 (2003)
  [Erratum-ibid.\  D {\bf 69}, 099901 (2004)]
  [arXiv:hep-ph/0210112].
\vspace{-0.3mm}

\bibitem{Barger:2006fm}
  V.~Barger, J.~Jiang, P.~Langacker and T.~Li,
  Int.\ J.\ Mod.\ Phys.\  A {\bf 22}, 6203 (2007)
  [arXiv:hep-ph/0612206].
\vspace{-0.3mm}

\bibitem{Babu:2008ge}
  K.~S.~Babu, I.~Gogoladze, M.~U.~Rehman and Q.~Shafi,
  Phys.\ Rev.\  D {\bf 78}, 055017 (2008)
  [arXiv:0807.3055 [hep-ph]].
\vspace{-0.3mm}

\bibitem{Liu:2009cc}
  C.~Liu,
  Phys.\ Rev.\  D {\bf 80}, 035004 (2009)
  [arXiv:0907.3011 [hep-ph]].
\vspace{-0.3mm}

\bibitem{Graham:2009gy}
  P.~W.~Graham, A.~Ismail, S.~Rajendran and P.~Saraswat,
  Phys.\ Rev.\  D {\bf 81}, 055016 (2010)
  [arXiv:0910.3020 [hep-ph]].
\vspace{-0.3mm}

\bibitem{Rosner:1985hx}
  J.~L.~Rosner,
  Comments Nucl.\ Part.\ Phys.\  {\bf 15}, 195 (1986).
\vspace{-0.3mm}

\bibitem{CDFtquark}
 T. Aaltonen et al., The CDF Collaboration, CDF Public Note March 5, 2010.
  (``Search for Heavy Top t'  to W q in Lepton Plus Jets Events in $4.6 {\rm fb}^{-1}$'').
\vspace{-0.3mm}

\bibitem{Aaltonen:2007je}
  T.~Aaltonen {\it et al.}  [CDF Collaboration],
  Phys.\ Rev.\  D {\bf 76}, 072006 (2007)
  [arXiv:0706.3264 [hep-ex]].
\vspace{-0.3mm}

  \bibitem{Isidori:2010kg}
  G.~Isidori, Y.~Nir and G.~Perez,
  arXiv:1002.0900 [hep-ph].
\vspace{-0.3mm}

\bibitem{Chivukula:1987py}
  R.~S.~Chivukula and H.~Georgi,
  Phys.\ Lett.\  B {\bf 188}, 99 (1987).
\vspace{-0.3mm}

  \bibitem{D'Ambrosio:2002ex}
  G.~D'Ambrosio, G.~F.~Giudice, G.~Isidori and A.~Strumia,
  Nucl.\ Phys.\  B {\bf 645}, 155 (2002)
  [arXiv:hep-ph/0207036].
\vspace{-0.3mm}

  \bibitem{Feldmann:2008ja}
  T.~Feldmann and T.~Mannel,
  Phys.\ Rev.\ Lett.\  {\bf 100}, 171601 (2008)
  [arXiv:0801.1802 [hep-ph]].
\vspace{-0.3mm}

  \bibitem{Kagan:2009bn}
  A.~L.~Kagan, G.~Perez, T.~Volansky and J.~Zupan,
  arXiv:0903.1794 [hep-ph].
\vspace{-0.3mm}

  \bibitem{Feldmann:2009dc}
  T.~Feldmann, M.~Jung and T.~Mannel,
  arXiv:0906.1523 [hep-ph].
\vspace{-0.3mm}

 \bibitem{Manohar}
  A.~V.~Manohar and M.~B.~Wise,
  Phys.\ Rev.\  D {\bf 74}, 035009 (2006)
  [arXiv:hep-ph/0606172].
\vspace{-0.3mm}

\bibitem{Kim:2008bx}
  C.~Kim and T.~Mehen,
  Phys.\ Rev.\  D {\bf 79}, 035011 (2009)
  [arXiv:0812.0307 [hep-ph]].
\vspace{-0.3mm}

\bibitem{Idilbi:2009cc}
  A.~Idilbi, C.~Kim and T.~Mehen,
  Phys.\ Rev.\  D {\bf 79}, 114016 (2009)
  [arXiv:0903.3668 [hep-ph]].
\vspace{-0.3mm}

\bibitem{Burgess:2009wm}
  C.~P.~Burgess, M.~Trott and S.~Zuberi,
  JHEP {\bf 0909}, 082 (2009)
  [arXiv:0907.2696 [hep-ph]].
\vspace{-0.3mm}

\bibitem{Fornal:2010ac}
  B.~Fornal and M.~Trott,
  JHEP {\bf 1006}, 110 (2010)
  [arXiv:1001.4287 [hep-ph]].
\vspace{-0.3mm}

\bibitem{Arnold:2009ay}
  J.~M.~Arnold, M.~Pospelov, M.~Trott and M.~B.~Wise,
  JHEP {\bf 1009}, 073 (2010)
  [arXiv:0911.2225 [hep-ph]].
\vspace{-0.3mm}

  \bibitem{Glashow:1976nt}
  S.~L.~Glashow and S.~Weinberg,
  Phys.\ Rev.\  D {\bf 15}, 1958 (1977).
\vspace{-0.3mm}

\bibitem{Grossman:2007bd}
  Y.~Grossman, Y.~Nir, J.~Thaler, T.~Volansky and J.~Zupan,
  Phys.\ Rev.\  D {\bf 76}, 096006 (2007)
  [arXiv:0706.1845 [hep-ph]].
\vspace{-0.3mm}

\bibitem{Gross:2010ce}
  E.~Gross, D.~Grossman, Y.~Nir and O.~Vitells,
  Phys.\ Rev.\  D {\bf 81}, 055013 (2010)
  [arXiv:1001.2883 [hep-ph]].
\vspace{-0.3mm}

\bibitem{Erler:2009jh}
  J.~Erler, P.~Langacker, S.~Munir and E.~R.~Pena,
  arXiv:0906.2435 [hep-ph].
\vspace{-0.3mm}

  \bibitem{CDFbquark}
 T. Aaltonen et al., The CDF Collaboration,
 Phys. Rev. Lett. {\bf 104}, 091801 (2010)
  [arXiv:0912.1057 [hep-ex]].
\vspace{-0.3mm}


\bibitem{Passarino:1978jh}
  G.~Passarino and M.~J.~G.~Veltman,
  Nucl.\ Phys.\  B {\bf 160}, 151 (1979).
\vspace{-0.3mm}

\bibitem{Holdom:1990tc}
  B.~Holdom and J.~Terning,
  Phys.\ Lett.\  B {\bf 247}, 88 (1990).
\vspace{-0.3mm}

 \bibitem{Peskin:1990zt}
  M.~E.~Peskin and T.~Takeuchi,
  Phys.\ Rev.\ Lett.\  {\bf 65}, 964 (1990).
\vspace{-0.3mm}

  \bibitem{Golden:1990ig}
  M.~Golden and L.~Randall,
  Nucl.\ Phys.\  B {\bf 361}, 3 (1991).
\vspace{-0.3mm}

  \bibitem{Maksymyk:1993zm}
  I.~Maksymyk, C.~P.~Burgess and D.~London,
  Phys.\ Rev.\  D {\bf 50}, 529 (1994)
  [arXiv:hep-ph/9306267].
\vspace{-0.3mm}
  \bibitem{Burgess:1993mg}
  C.~P.~Burgess, S.~Godfrey, H.~Konig, D.~London and I.~Maksymyk,
  Phys.\ Lett.\  B {\bf 326}, 276 (1994)
  [arXiv:hep-ph/9307337].
\vspace{-0.3mm}
\bibitem{Gresham:2007ri}
  M.~I.~Gresham and M.~B.~Wise,
  Phys.\ Rev.\  D {\bf 76}, 075003 (2007)
  [arXiv:0706.0909 [hep-ph]].
\vspace{-0.3mm}
  \bibitem{Amsler:2008zzb}
  C.~Amsler {\it et al.}  [Particle Data Group],
  Phys.\ Lett.\  B {\bf 667}, 1 (2008).
\vspace{-0.3mm}
\bibitem{Bamert:1996px}
  P.~Bamert, C.~P.~Burgess, J.~M.~Cline, D.~London and E.~Nardi,
  Phys.\ Rev.\  D {\bf 54}, 4275 (1996)
  [arXiv:hep-ph/9602438].
\vspace{-0.3mm}
\bibitem{Barger:1995dd}
  V.~D.~Barger, M.~S.~Berger and R.~J.~N.~Phillips,
  Phys.\ Rev.\  D {\bf 52}, 1663 (1995)
  [arXiv:hep-ph/9503204].
\vspace{-0.3mm}
\bibitem{Inami:1980fz}
  T.~Inami and C.~S.~Lim,
  Prog.\ Theor.\ Phys.\  {\bf 65}, 297 (1981)
  [Erratum-ibid.\  {\bf 65}, 1772 (1981)].
\vspace{-0.3mm}
\bibitem{Buchalla:1995vs}
  G.~Buchalla, A.~J.~Buras and M.~E.~Lautenbacher,
  Rev.\ Mod.\ Phys.\  {\bf 68}, 1125 (1996)
  [arXiv:hep-ph/9512380].
\vspace{-0.3mm}
\bibitem{Bjorken:2002vt}
  J.~D.~Bjorken, S.~Pakvasa and S.~F.~Tuan,
  Phys.\ Rev.\  D {\bf 66}, 053008 (2002)
  [arXiv:hep-ph/0206116].
\vspace{-0.3mm}
\bibitem{Golowich:2007ka}
  E.Golowich, J.Hewett, S.Pakvasa and A.A.Petrov,
  Phys.Rev.\  D {\bf 76}, 095009 (2007)
  [arXiv:0705.3650 [hep-ph]].
\vspace{-0.3mm}

\bibitem{Artuso:2005ym}
  M.~Artuso {\it et al.}  [CLEO Collaboration],
  Phys.\ Rev.\ Lett.\  {\bf 95}, 251801 (2005)
  [arXiv:hep-ex/0508057].
\vspace{-0.3mm}
\bibitem{Gupta:1996yt}
  R.~Gupta, T.~Bhattacharya and S.~R.~Sharpe,
  Phys.\ Rev.\  D {\bf 55}, 4036 (1997)
  [arXiv:hep-lat/9611023].
\vspace{-0.3mm}
\bibitem{Ciuchini:1997bw}
  M.~Ciuchini, E.~Franco, V.~Lubicz, G.~Martinelli, I.~Scimemi and L.~Silvestrini,
  Nucl.\ Phys.\  B {\bf 523}, 501 (1998)
  [arXiv:hep-ph/9711402].
\vspace{-0.3mm}
\bibitem{Abe:2007rd}
  K.~Abe {\it et al.}  [BELLE Collaboration],
  Phys.\ Rev.\ Lett.\  {\bf 99}, 131803 (2007)
  [arXiv:0704.1000 [hep-ex]].
\vspace{-0.3mm}
\bibitem{Donoghue:1985hh}
  J.~F.~Donoghue, E.~Golowich, B.~R.~Holstein and J.~Trampetic,
  Phys.\ Rev.\  D {\bf 33}, 179 (1986).
\vspace{-0.3mm}

\bibitem{Herrlich:1996vf}
  S.~Herrlich and U.~Nierste,
  Nucl.\ Phys.\  B {\bf 476}, 27 (1996)
  [arXiv:hep-ph/9604330].
\vspace{-0.3mm}

\bibitem{Lubicz:2008am}
  V.~Lubicz and C.~Tarantino,
  Nuovo Cim.\  {\bf 123B}, 674 (2008)
  [arXiv:0807.4605 [hep-lat]].
\vspace{-0.3mm}

\bibitem{Grinstein:1987vj}
  B.~Grinstein, R.~P.~Springer and M.~B.~Wise,
  Phys.\ Lett.\  B {\bf 202}, 138 (1988).
\vspace{-0.3mm}

  \bibitem{Grzadkowski:2008mf}
  B.~Grzadkowski and M.~Misiak,
  Phys.\ Rev.\  D {\bf 78}, 077501 (2008)
  [arXiv:0802.1413 [hep-ph]].
\vspace{-0.3mm}
  \bibitem{Barberio:2007cr}
  E.~Barberio {\it et al.}  [Heavy Flavor Averaging Group (HFAG)
                  Collaboration],
  arXiv:0704.3575 [hep-ex].
\vspace{-0.3mm}
  \bibitem{Barger:1985nq}
  V.~D.~Barger, N.~Deshpande, R.~J.~N.~Phillips and K.~Whisnant,
  Phys.\ Rev.\  D {\bf 33}, 1912 (1986)
  [Erratum-ibid.\  D {\bf 35}, 1741 (1987)].
\vspace{-0.3mm}
\bibitem{AguilarSaavedra:2005pv}
  J.~A.~Aguilar-Saavedra,
  Phys.\ Lett.\  B {\bf 625}, 234 (2005)
  [Erratum-ibid.\  B {\bf 633}, 792 (2006)]
  [arXiv:hep-ph/0506187].
\vspace{-0.3mm}

\bibitem{Andre:2003wc}
  T.~C.~Andre and J.~L.~Rosner,
  Phys.\ Rev.\  D {\bf 69}, 035009 (2004)
  [arXiv:hep-ph/0309254].
\vspace{-0.3mm}
\bibitem{AguilarSaavedra:2009es}
  J.~A.~Aguilar-Saavedra,
  JHEP {\bf 0911}, 030 (2009)
  [arXiv:0907.3155 [hep-ph]].
\vspace{-0.3mm}
\bibitem{Berger:2009qy}
  E.~L.~Berger and Q.~H.~Cao,
  Phys.\ Rev.\  D {\bf 81}, 035006 (2010)
  [arXiv:0909.3555 [hep-ph]].
\vspace{-0.3mm}

  \bibitem{Skiba:2007fw}
  W.~Skiba and D.~Tucker-Smith,
  Phys.\ Rev.\  D {\bf 75}, 115010 (2007)
  [arXiv:hep-ph/0701247].
\vspace{-0.3mm}
\bibitem{Holdom:2007nw}
  B.~Holdom,
  JHEP {\bf 0703}, 063 (2007)
  [arXiv:hep-ph/0702037].
\vspace{-0.3mm}

\bibitem{Czakon:2008ii}
  M.~Czakon and A.~Mitov,
  Nucl.\ Phys.\  B {\bf 824}, 111 (2010)
  [arXiv:0811.4119 [hep-ph]].

\vspace{-0.3mm}

\bibitem{Martin:2009iq}
  A.D.Martin, W.J.Stirling, R.S.Thorne and G.Watt,
  Eur.\ Phys.\ J.\  C {\bf 63}, 189 (2009)
  [arXiv:0901.0002 [hep-ph]].
\vspace{-0.3mm}

\bibitem{vanRitbergen:1997va}
  T.~van Ritbergen, J.~A.~M.~Vermaseren and S.~A.~Larin,
  Phys.\ Lett.\  B {\bf 400}, 379 (1997)
  [arXiv:hep-ph/9701390].
\vspace{-0.3mm}
\bibitem{Abbate:2010vw}
  R.~Abbate, M.~Fickinger, A.~Hoang, V.~Mateu and I.~W.~Stewart,
  arXiv:1004.4894 [hep-ph].
\vspace{-0.3mm}


\end{thebibliography}
\end{document}